%% file: main.tex
\documentclass[10pt,journal,compsoc]{IEEEtran}

\usepackage{enumerate}
\usepackage{graphicx}
\usepackage{xcolor}
\usepackage{ulem}
\usepackage{multirow}
\usepackage{hyperref}
\usepackage{booktabs}
\usepackage{wrapfig}

\ifCLASSOPTIONcompsoc
  \usepackage[nocompress]{cite}
\else
  \usepackage{cite}
\fi

\def\WrongDPOAdvice{\textsc{Wrong-Advice}}
\def\IgnoredDPOAdvice{\textsc{Ignored-DPO-Advice}}
\def\DPOAdviceNotSought{\textsc{DPO-Advice-Not-Sought}}
\def\WebsiteForcesChoices{\textsc{Website-Forces-Choices}}
\def\AdminAsksForEverything{\textsc{Admin-Asks-For-Everything}}
\def\NeglectedSubjectRight{\textsc{Neglected-Subject-Right}}
\def\SubjectRightRequest{\textsc{Subject-Right-Request}}
\def\NoDataProtectionPrinciples{\textsc{No-Data-Protection-Principles}}
\def\UncheckedRemoteMonitoring{\textsc{Unchecked-Remote-Monitoring}}
\def\WrongPublicProcurement{\textsc{Wrong-Public-Procurement}}
\def\SoftwareEndofLife{\textsc{Software-End-of-Life}}
\def\SubcontractorViolatesPrivacy{\textsc{Subcontractor-Violates-Privacy}}

\newenvironment{bio}[1]
{\par
 \bigskip
 \begin{wrapfigure}{l}[0pt]{1in}
 \vspace{-15pt}
 \includegraphics[height=1in,clip,keepaspectratio]{#1}
 \vspace{-15pt}
 \end{wrapfigure}
 \footnotesize \noindent}
{\par\bigskip}

\begin{document}

\onecolumn

\begin{figure}
    \centering
    \includegraphics[width=.3\textwidth]{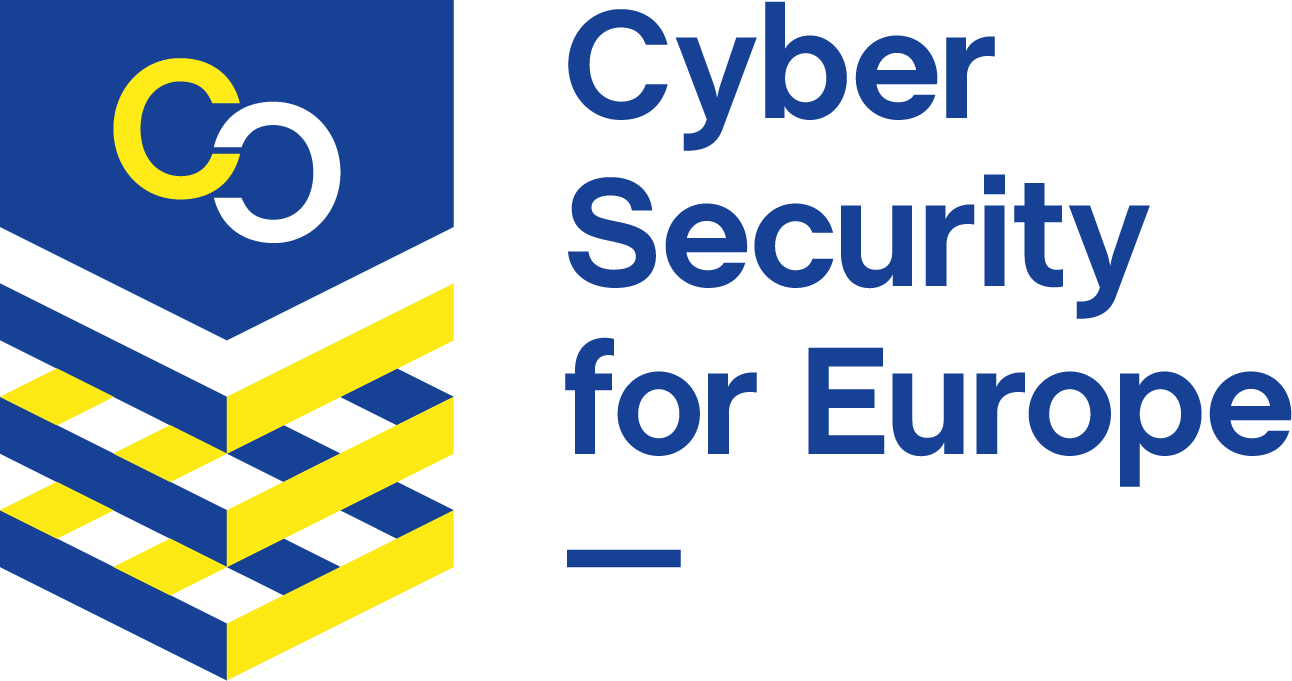}
\end{figure}

\vspace{2\baselineskip}

\begin{center}
{\huge \textbf{The Data Protection Officer, \\ an ubiquitous role nobody really knows}}
\end{center}

\vspace{20\baselineskip}

{\large
Authors:
\begin{itemize}
    \item[] \textbf{Francesco Ciclosi}, University of Trento (IT), on leave from the Italian Ministry of Economic Development (IT)
    \item[] \textbf{Fabio Massacci}, University of Trento (IT), Vrije Universiteit Amsterdam (NL)
\end{itemize}
}

\vfill

\begin{wrapfigure}{l}{2.5cm}
\vspace{-\baselineskip}
\includegraphics[width=2.5cm]{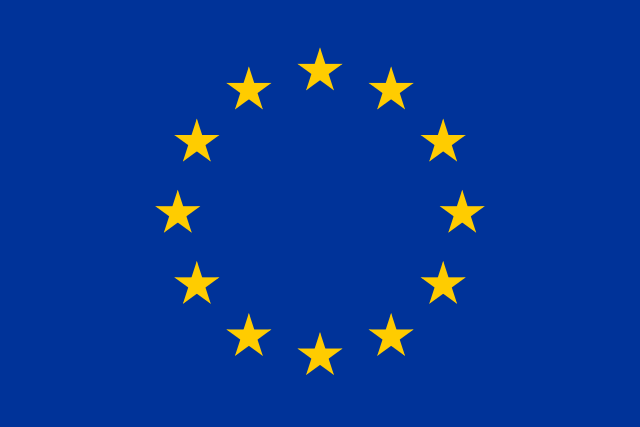}
\end{wrapfigure}

This paper was written within the H2020 CyberSec4Europe project that received funding from the European Union's Horizon 2020 research and innovation programme under grant agreement No 830929. This paper reflects only the author's view and the Commission is not responsible for any use that may be made of the information contained therein.

\clearpage
\twocolumn

\begin{bio}{CSfE_Logo_Reverse}
\textbf{Cyber Security for Europe (CyberSec4Europe)} As a research project, CyberSec4Europe is working towards harmonising the journey from the development of software components that fit the requirements identified by a set of short- and long-term roadmaps, leading to a series of consequent recommendations. These are tied to the project’s real-world demonstration use cases that address cybersecurity challenges within the vertical sectors of digital infrastructure, finance, government and smart cities, healthcare and transportation.
CyberSec4Europe’s long-term goal and vision are of a European Union that has all the capabilities required to secure and maintain a healthy democratic society, living according to European constitutional values, with regard to, for example, privacy and data sharing, and being a world-leading digital economy.
CyberSec4Europe’s main objective is to pilot the consolidation and future projection of the cybersecurity capabilities required to secure and maintain European democracy and the integrity of the Digital Single Market. CyberSec4Europe has translated this broad objective into measurable, concrete steps: three policy objectives, three technical objectives and two innovation objectives. CyberSec4Europe: \textbf{Cyber Security for Europe}. More information at \textbf{\url{https://cybersec4europe.eu/}}.
\end{bio}

\begin{bio}{fciclosi}
\textbf{Francesco Ciclosi} is a Ph.D. student in information engineering and computer science at the University of Trento, Italy. He has been a DPO for an Italian city for the past five years and is currently on leave from the Italian Ministry of Economic Development. He works on methodologies for privacy in socio-technical systems. Previously, he was an adjunct professor of computer science at the University of Macerata. Contact him at \emph{francesco.ciclosi@unitn.it}.
\end{bio}

\vspace*{-\baselineskip}

\begin{bio}{Fabio-Massacci-LowRes}
\textbf{Fabio Massacci} (Phd 1997) is a professor at the University of 
Trento, Italy, and Vrije Universiteit Amsterdam, The Netherlands. 
He received the Ten Years Most Influential Paper award by the IEEE Requirements Engineering Conference in 2015. 
He is the Leader of Education and Skill WP of the CyberSec4Europe project. Contact him at \emph{fabio.massacci@ieee.org}.
\end{bio}

How to cite this paper:
\begin{itemize}
    \item Ciclosi F. and Massacci F., The Data Protection Officer, an ubiquitous role nobody really knows. \emph{IEEE Security \& Privacy, Special Issue on Usable Security for Security Workers}. IEEE Press. 2023, doi: 10.1109/MSEC.2022.3222115. Full arXiv version with supplemental material.
\end{itemize}

License:
\begin{itemize}
\item This article is made available with a perpetual, non-exclusive, non-commercial license to distribute.
\end{itemize}

\clearpage

\twocolumn \label{main:paper}

\title{The Data Protection Officer, \\ an ubiquitous role nobody really knows}

\author{Francesco~Ciclosi,~\IEEEmembership{}
       Fabio~Massacci,~\IEEEmembership{Member,~IEEE,}

\IEEEcompsocitemizethanks{\IEEEcompsocthanksitem F. Ciclosi (corresponding author) is with the University of Trento, Italy, and on leave from the Italian Ministry of Economic Development.\protect\\
E-mail: francesco.ciclosi@unitn.it
\IEEEcompsocthanksitem F. Massacci is with the University of Trento, Italy and Vrije Universiteit Amsterdam, The Netherlands.}

\thanks{Manuscript accepted 02/11/2022}
}

\markboth{Accepted for IEEE Security and Privacy magazine}{Ciclosi and Massacci}

\IEEEtitleabstractindextext{%
\begin{abstract}
Among all cybersecurity and privacy workers, the Data Protection Officer (DPO) stands between those auditing a company's compliance and those acting as management advisors. A person that must be somehow versed in legal, management, and cybersecurity technical skills. We describe how this role tackles socio-technical risks in everyday scenarios.
\end{abstract}

}

\maketitle

\IEEEdisplaynontitleabstractindextext

\IEEEpeerreviewmaketitle

\IEEEraisesectionheading{\section{Introduction}\label{sec:introduction}}

\IEEEPARstart{T}{he} recent application of Regulation (EU) 2016/679, best known to the world as the \emph{General Data Protection Regulation} or GDPR, introduced the role of the data protection officer (DPO). While DPOs have been a key enabler of the GDPR \cite{AlmeidaTeixeira2019}, the role of this privacy worker is not a new concept: in several EU Member States, its appointment was already good practice for some years.
Yet, the GDPR does not formally describe the DPO job profile, and most papers discuss how to support a DPO with algorithms without providing practical examples of what the DPO does.

For example, Diamantopoulou et al.\cite{Diamantopoulou2020} identify which ISO 27001/2 controls need to be extended to meet GDPR requirements and which of them the DPO is involved - but why in some and not in others? Ryan et al. \cite{ryan2020} explain how their framework RegTech can be helpful to a DPO for checking GDPR compliance - but when and for what concretely? Chatzipolidis et al. \cite{CHATZIPOULIDIS2019} describe a readiness assessment tool for GDPR's compliance - but for solving social or technical issues? Other articles discuss GDPR’s compliance topics as if DPOs did not exist, from software engineering \cite{Martin2018} to socio-technical management processes \cite{MALATJI2020}, from the GDPR's cost among cybersecurity investments \cite{Layton2019}, to algorithms for checking GDPR compliance itself\cite{Basin2018}.

Our purpose is to introduce this legally required organizational role --- this ubiquitous privacy worker --- to the engineering community represented by Security \& Privacy readers through concrete examples of what problems DPOs face, what they do, and what they may or must know. While the literature is surprisingly silent about this, we think that the knowledge of the everyday challenges that DPOs have is the starting point for all subsequent research activities. For example, if a researcher has no reference to the daily activities of \emph{the} key privacy worker in charge of GDPR compliance, how can one design logic or a tool for checking this privacy compliance or any privacy-by-design technology with a practical impact? 
In summary, our key research question is: 
\begin{itemize}
\item \textit{Can we enucleate in a few representative scenarios the concrete activities of a DPO?}
\end{itemize} 

The article focuses on the role of the DPO introduced by the GDPR, but the insights are valuable for readers outside the EU countries. The GDPR can apply to organizations that carry out their activities in the EU and organizations outside the EU that process the personal data of EU data subjects. Further, in many countries worldwide, there is data protection legislation in which a DPO role exists at some level. The International Association of Privacy Professionals (IAPP) lists the different roles in many countries worldwide that share some characteristics with the DPO legally defined in the EU \cite{IAPP2021}.

\section{Our Methodology}\label{sec:methodology}
The insights described in this article are grounded in case studies along Yin's case study methodology \cite[Ch.4]{yin2018case} and the suggestions \cite{glaser2004remodeling} by Glaser, the founder of grounded theory, to build core categories across field observations derived from the live experience.

First, we analyzed data protection laws and recommendations of relevant authorities. Secondly, we analyzed the seven functions of the DPO that the European Data Protection Supervisor (EDPS) identified in its paper on the role of DPO in compliance with Regulation (EC) 45/2001. Then, we looked at the summary of opinions of some supervisory authorities (i.e., Bulgaria, Croatia, Italy, Poland, and Spain) on the DPO’s activities involved with these functions (e.g., \cite{Korff2019}) to have a perspective that was not restricted to a single country.

To make this paper concrete as a use cases references, we selected only sources of information for which there was evidence that the activities carried out by DPO involve at least one of these seven functions. The starting point for the case study selection was the personal experience of the first author, who has been a DPO in the Italian public administration for the past five years and is a member of the Italian Association of DPOs.

To make the results of our study accessible, we looked for some publicly available information (for example, court decisions, supervisory authorities' decisions, and newspaper articles) on case studies similar to the ones on which the first author had first-hand experience. This approach allows us to go beyond the individual experience and provide shareable evidence. Out of over 90 public case studies, we finally distilled 12 scenarios with at least one analyzable practical example for each of the DPO's primary functions. In this paper's main text, we made the scenario general by renaming the actors involved (but without altering their nature) to be resilient to potential requests on the "right to be forgotten" and be of general interest to the readers.

The scenarios we propose map well into the general literature. For example, the site \textit{GDPR Enforcement Tracker fines database} \cite{EnforcementTracker} reports 1475 occurrences of GDPR fines related to DPOs across 31 different EU Member or EEA States. Our scenarios cover more than 1400 cases.

The concrete references 
are listed in the supplemental material available in Appendix \ref{Appendix_Case_studies_references}.

\section{How this role is born}\label{sec:RoleBorn}
\begin{figure*}[!t]
\centering
\includegraphics[scale=0.56]{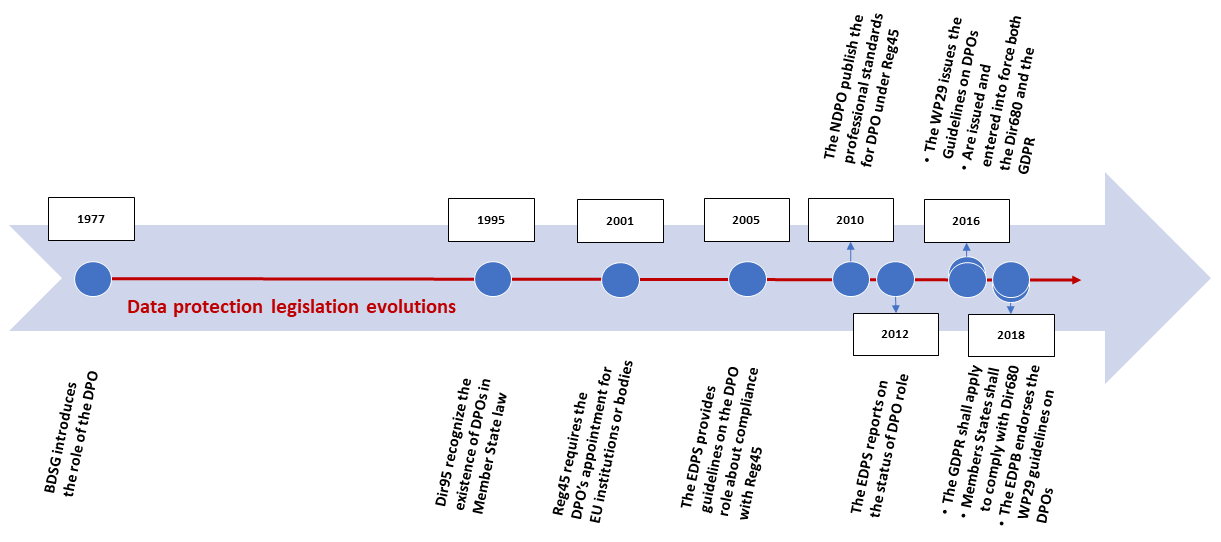}
\caption{The evolution of the European role of Data Protection Officer (DPO)}
\label{fig:DPOstory}
\end{figure*}

In Europe, the concept of DPO comes from the German data protection law of 1977, the Bundesdatenschutzgesetz (BDSG), which introduced a precursor of the role (Figure \ref{fig:DPOstory}).

The DPO role over time became widely adopted by the other European States until, in 1995, the European Community issued Directive 95/46/EC on the protection of individuals concerning the processing of personal data and on the free movement of such data. A patchwork of approaches followed: many Member States introduced the DPO role in their national law, as in Austria (where the appointment was mandatory) or in France (where the appointment was optional), but only in some of them (in Italy, the DPO role was absent in national law). Moreover, the DPO duties were limited to independently ensuring an organization's internal application of the national provisions taken according to the Directive and keeping a register of processing operations carried out by the controller.

For EU institutions, only the appointment of at least one DPO was mandated by Regulation (EC) 45/2001. These rules were very similar to the ones that would be introduced in later years.
2016 was a pivotal year for data protection as the European Parliament and the Council issued the Directive (EU) 2016/680 on criminal offenses or criminal penalties and the GDPR on personal data protection.

To provide an interpretation of the EU's data protection legislation, the Article 29 Working Party Committee issued the Guidelines on DPOs (WP243) \cite{WP243rev01}, which was initially adopted on December 13, 2016, and later revised on April 5, 2017. After the GDPR adoption, the new European Data Protection Board (EDPB), an independent European body tasked with ensuring the consistent application of data protection rules throughout the European Union, endorsed these guidelines in its first plenary meeting on May 25, 2018.

Not all organizations are required to appoint a DPO, although doing it is a good practice. There are three specific cases in which a controller or a processor must appoint a DPO (GDPR article 37). In the first case, if the organization is a public institution, it must appoint a DPO. Otherwise, an organization must appoint a DPO if it requires regular and systematic monitoring of data subjects on a large scale (second case) or processes on a large scale personal data which belong to special categories or are related to criminal offenses (third case).

The GDPR Article 39 entrusts to DPO different tasks:
\begin{enumerate}[a)]
    \item to inform and advise one's employer on how to carry out processing about its obligations under the law;
    \item to monitor compliance with the law;
    \item to provide advice regards the data protection impact assessment (DPIA) and monitoring its execution;
    \item to cooperate with the supervisory authority (SA) as a contact point between this and the organization.
\end{enumerate}

\begin{table*}
\centering
\caption{High-Level View of the Functions of the DPOs}\label{tab:function}
\begin{minipage}{\textwidth}\footnotesize
This table describes the tasks of the DPO grouped according to the seven functions of the DPO that the European Data Protection Supervisor (EDPS) identified in its position paper on the role of DPO in compliance with Regulation (EC) 45/2001.
\end{minipage}
\begin{tabular}{p{0.3\textwidth}p{0.7\textwidth}}
\toprule
DPO’s Functions & 
Summary Descriptions \\
\midrule
Organizational function & %
Review or even directly organize a processing operations register on behalf of the controller, help both assess the related risks, and support the processing activities with high-risk value. \\
Monitoring of compliance & 
Investigate (on autonomous initiative) matters and occurrences directly relating to the GDPR and report back to the controller \\
Advisory function &
Make recommendations for the practical improvement of data protection to the controller and advise it on matters concerning the related provisions.\\
Cooperative function &
Facilitate cooperation (between the Supervisory Authority and the controller), especially in the frame of investigations, complaint handling, or prior checks. \\
Handle queries or complaints &
The authorization to handle queries or complaints originated from the very possibility of autonomous investigations. \\
Information and raising awareness function &
Prepare staff information notes, training sessions, privacy statements, and learning material. \\
Enforcement &
Powers of enforcement are limited. \\
\bottomrule
\end{tabular}
\end{table*}

Table \ref{tab:function} summarizes the seven functions of the DPO, identified by the EDPS in its position paper on this role.
Appendix \ref{sec:DPOtasks} provides further information and examples about these seven functions of the DPO. 

In summary, DPOs must be fully cognizant of the controller's working environment to carry out their tasks. This awareness implies that a DPO must know the internal distribution and allocation of responsibilities and tasks related to every personal data processing. A DPO must also be familiar with any external links (between the controller and other organizations) and with legal frameworks in which these links take place. According to many supervisory authorities' opinions \cite{Korff2019}, a preliminary task in which a DPO scopes the controller’s environment fulfills this requirement.

\section{What problems does a DPO face?}\label{sec:DPOscenarios}

To understand the daily problems a DPO faces, we have illustrated several real-life scenarios in Table \ref{tab:scenarios}. We edited them to obfuscate the original entity responsible for the privacy issues faced by the DPO. The supplemental material reports the sources of these scenarios.

Empirical research on the problems of the DPO \cite{casutt2020data} has shown that sometimes these could be very basic and relate to a lack of sufficient resources (time, finances, and humans) to carry out one's duties and to some issues in the operational interpretation of the law.
Below in Table \ref{tab:scenarios}, we analyze the challenges that DPOs face even when adequately supported.

\begin{table*}
\centering 
\caption{Scenarios and Privacy Issues daily faced by a DPO}\label{tab:scenarios}
\footnotesize
\begin{tabular}{p{0.18\textwidth}p{0.37\textwidth}p{0.37\textwidth}}
\toprule
\textbf{Short name} & 
\textbf{Scenario description} & 
\textbf{What went wrong} \\
\midrule

\WrongDPOAdvice & A controller wants to process data using a processor service and asks the DPO’s advice to understand if the proposed contract complies with GDPR.  & The advice of the DPO turned out to be wrong, but the controller uncritically trusted the DPO's advice.\\
\hline

\IgnoredDPOAdvice & A controller implements a new data processing. The DPO advises that a prior execution of a DPIA is required. & The controller chooses to neglect the DPO’s advice without justifying in writing why it did not take into account that advice. \\ 
\hline

\DPOAdviceNotSought & A controller implements a registration procedure of service without prior asking for DPO’s advice about compliance with GDPR principles & The registration procedure does not carry out a check on the identity of the person who enrolls, so it is unknown who saw the data\\ 
\hline

\WebsiteForcesChoices & In a controller procedure, data processing for marketing purposes asks for the consent of the data subject. & The procedure forces a data subject to release consent, imposing him or her to select a box, otherwise preventing it from continuing. \\
\hline

\AdminAsksForEverything & A controller’s staff member satisfies the law on access to data (e.g., FOIA) by publishing some documents about a data subject. & A controller processes personal data by asking the data subject and publishing all data without correctly applying the data minimization principle. \\ 
\hline

\NeglectedSubjectRight & A data subject files a complaint with the competent supervisory authority because s/he has not received a response within the time frame set by law & The controller did not designate the DPO, or it did, but the designated DPO did not monitor the official address. \\ 
\hline

\SubjectRightRequest & A data subject exercises one's rights of access according to article 15 of the GDPR, sending a formal request to the DPO’s address. & When the controller drew up the record of processing activities, it did not correctly identify the actual processing activities, so the DPO could not answer the query. \\ 
\hline

\NoDataProtectionPrinciples & A controller implements a configuration in company equipment. & There is an incorrect configuration that allows unfettered access to personal data. \\ 
\hline

\UncheckedRemoteMonitoring & A controller implements remote assistance software tools on company workstations & The technicians use a remote assistance software solution that does not notify the user when remote access is performed. \\
\hline 

\WrongPublicProcurement & A controller issue a public tender for procuring products or services. & In the tender evaluation grid, there is no explicit checkpoint for applicants to "demonstrate" that their products or services fully comply with the GDPR. \\ 
\hline

\SoftwareEndofLife & The DPO of a company noticed some results about the software used for processing activities. & The software used in the company's workstations is obsolete, and the support provided by the vendor software house is expired or close to expiring. \\ 
\hline

\SubcontractorViolatesPrivacy & A controller appoints its DPO to test a software procedure of a tender's winner & The tender winner has violated the contract for the supply of IT programs and services terms on GDPR compliance. \\

\bottomrule
\end{tabular}
\end{table*}

A first example: although maintaining a record of processing activities is formally a controller's duty, the DPO will most likely be in charge of this work or closely involved in its oversight activities. In quite some job adverts for DPO appointments, it is formally stated that the DPO is in charge of maintaining the record of processing activities. 
Without a regulatory constraint, the DPO may also be appointed to carry out some activities that are formally a duty of the controller. It is a free choice of the controller who pays the DPO. The involvement of the DPO is only sometimes an indicator that things are working well. A data controller could use a non-GDPR-compliant service because of a DPO’s mistake (Tab. \ref{tab:scenarios}, \WrongDPOAdvice). Unfortunately, the controller is solely responsible for the choice made, and even a DPO's evaluation errors expose it to administrative fines or penalties. The accountability principle constrains the controller to demonstrate having complied with the regulatory requirement.

In an opposite scenario, a controller operating a catering service implements a new data processing to control the EU Digital Covid Certificate of staff but neglects the DPO’s advice without justifying in writing why that advice has not been taken into account (\IgnoredDPOAdvice). This action is to blame because it is reasonable that the DPO gave his or her advice to ensure that the processing complied with the GDPR. Again, not documenting the reasons behind choosing to neglect this advice results in violating the accountability principle, exposing data subjects to risks and the controller itself to administrative fines or penalties.

The DPIA execution is a controller’s responsibility, while the DPO task is limited to providing the advice requested. The WP29 \cite{WP243rev01} highlights that a controller should justify in black and white in the DPIA's documentation why it has not considered the DPO's advice.
For properly handling the execution of a DPIA, a controller could define, in an internal regulation, the procedures for consulting the DPO about this topic. By way of example, this regulation may contain whether or not to carry out a DPIA, what methodology to follow, and whether to use internal resources or outsource it. Other helpful information to include in regulation is what safeguards to apply to mitigate any risks to the rights and freedoms of the data subjects.

The scenario \DPOAdviceNotSought\ describes a case in which a controller is exposed to mistakes in the processing's design because it did not ask the DPO's advice.
The scenario \WebsiteForcesChoices\ exemplifies the vulnerability stemming from a wrong implementation of a seat reservation software procedure of a transport company. The software forced the data subject to consent to other forms of processing.

The \AdminAsksForEverything\ scenario is related to a failed application of the minimization principle. In it, the human resources office of a public institution (which must satisfy the law on publication obligation) publishes in the \textit{“Transparent Administration”} section of the institution's website the unredacted CV of the winner of a public selection, thus exposing the data subject’s personal data (e.g., home address, personal phone number, and others).

The relationship between the DPO and the supervisory authority (SA) is significant. The WP29 \cite{WP243rev01} highlights that the DPO must act as a “\emph{facilitator}” by cooperating with the SA. Furthermore, the obligation of secrecy or confidentiality cannot prohibit the DPO from contacting and seeking advice from the SA. The controller who acts to weaken this relationship could be sanctioned. If the data controller does not appoint the DPO, problems will likely arise in scenarios \NeglectedSubjectRight\ and \SubjectRightRequest. In \NeglectedSubjectRight, another violation is the missing communication to the SA of the DPO’s contact details because, in that case, the SA does not know how to contact the DPO to handle the complaint. In \SubjectRightRequest, an aggravating factor would occur if, when the controller draws up the record of processing activities, it does not correctly identify them. In such a case, even if appointed, the DPO may find it challenging to identify the processing details and promptly respond to the data subject's requests.

\section{Risk management}\label{sec:impactsanalysis}

DPOs face many challenges that we can classify into two categories: technical risks and socio-organizational risks. The first challenges stem from faulty technical or technological solutions which do not fully guarantee the protection of the subjects whose personal data is processed. The second type of challenge is due to incorrect organizational procedures or incorrect human behavior.  

We can subdivide technical risks into two subgroups. The first type relates to design problems when the DPO supports and advises the controller on technical choices. For example, the DPO may be asked to choose between configurations that comply with the data protection by default principle and configurations that seem to do so (Tab. \ref{tab:scenarios}, \NoDataProtectionPrinciples\ and \UncheckedRemoteMonitoring). S/he might find (or fail to find) insecure technical solutions in a call for tenders (\WrongPublicProcurement) that can cause kick-start litigation between contractors. The second type of risk involves misconfigurations or errors which appear during the implementation (\SubcontractorViolatesPrivacy\ and \SoftwareEndofLife).

Socio-organizational risk can appear daily while the DPO supports the controller in processing activities. We can divide them into four additional categories, such as auditing (Tab. \ref{tab:scenarios}, \IgnoredDPOAdvice\ and \SubcontractorViolatesPrivacy), communication (\NoDataProtectionPrinciples\ and \UncheckedRemoteMonitoring), processing designing (\SoftwareEndofLife) and relationship (\NeglectedSubjectRight).
Conflicting requirements are a particular instance of these risks. 

No matter how sophisticated the technical protection measures are, DPOs may experience side-channel attacks from the most unexpected human behavior. A well-designed processing activity may become unlawful because a staff member operates differently from what is prescribed by the procedure and from the instructions received by the controller. For example, an operator recorded the screen of the closed-circuit video surveillance cameras, only accessible to the local police control station, with a smartphone and disseminated a video of a traffic accident on social channels. 
Appendix \ref{sec:DPOchallenges} provides further information about technical and socio-organizational risks.

Table \ref{tab:vulnerabilitiesimpact} shows some of the possible consequences of damaging the freedom and rights of natural persons. It is difficult to determine the impact grade without an in-depth analysis of the processing and the technical and organizational means used to mitigate their risks. In some circumstances, the case study's high-level description (e.g., in the case of the GDPR accountability principle’s violation) is sufficient to conclude that the consequences are actual and potentially disastrous for an organization. Because personal data breach is a broad concept, we split this class into three classes representing the specific breach. Finally, Table \ref{tab:vulnerabilitiesimpact} includes some more specific classes (e.g., "\textit{GDPR non-compliant processing}" is a particular case of "\textit{unlawful processing}").

\begin{table*}
\centering
\caption{Some possible consequences of risks faced by a DPO may deal with}\label{tab:vulnerabilitiesimpact}
\begin{minipage}{\textwidth}\footnotesize
This table summarizes some examples of the consequences of technical or socio-technical risks that a DPO may face while carrying out his or her duties. These consequences are grouped into classes and associated with a possible reference scenario. Any such infringement exposes the controller to administrative fines of up to 20\,000\,000 EUR, or up to 4\% of the total worldwide annual turnover, whichever is higher. 
\end{minipage}
\begin{tabular}{p{0.15\textwidth}p{0.55\textwidth}p{0.25\textwidth}}
\toprule
\textbf{Impact's class} & \textbf{Impact example} & \textbf{Scenario example} \\
\midrule

Attack on customers & Identity theft of the data subject could happen (e.g., an attacker steals from a misconfigured database a data subject’s personal data and later uses them pretending to be the data subject). & \NoDataProtectionPrinciples, \DPOAdviceNotSought, \IgnoredDPOAdvice \\ 
\hline

Attack on employees & Unauthorized persons could capture the private data of employees & \UncheckedRemoteMonitoring\ \\ 
\hline

Contract termination & It is possible that the tender be subject to litigation due to a violation of the terms indicated in the contract for the supply of IT programs and services & \SubcontractorViolatesPrivacy\ \\ 
\hline

Data breach (access or disclosure) & A personal data breach may happen because unauthorized access to (or disclosure of) personal data transmitted, stored, or otherwise processed (e.g., a software’s misconfiguration allows technicians to connect to a workstation without the user's consent stealthily). & \UncheckedRemoteMonitoring, \SubcontractorViolatesPrivacy, \NoDataProtectionPrinciples \\
\hline

Data breach (destruction or loss ) & There is unlawful destruction or an accidental loss of personal data held by the controller (e.g., a malware of type ransomware has ciphered all office's files stored in a file server) & \SoftwareEndofLife \\ 
\hline

Data breach (alteration) & There is an unlawful alteration to personal data stored (e.g., exploiting a system’s vulnerability, a cracker modified some personal data stored in a database) & \SoftwareEndofLife, \NoDataProtectionPrinciples \\ 
\hline

Data Disclosure & Excess and irrelevant personal data are disseminated over the Internet by the controller. & \AdminAsksForEverything, \NeglectedSubjectRight\ \\ 
\hline

Inadequate identification & Anyone could pretend to be another data subject and access his or her personal data. & \DPOAdviceNotSought\ \\ 
\hline

GDPR non-compliant processing & A controller processes personal data without providing information to the data subject (e.g., a controller issues a loyalty card to a customer without providing the customer with a privacy policy) & \WrongDPOAdvice, \IgnoredDPOAdvice \\ 
\hline

Risky processing & The processing could be risky for the rights and freedoms of natural persons. & \IgnoredDPOAdvice\ \\ 
\hline
Unlawful processing & There is no legal basis for the processing. For example, the procedure forces the data subject to release consent for an unnecessary activity. In another example, a controller acquires personal data for a purpose and uses them for another purpose without obtaining the data subject's consent. & \WebsiteForcesChoices \\ 
\hline
Unlawful procurement & A contractor that offers products or services, not GDPR-compliant can win a bid. The call for tender may be subject to litigation. & \WrongPublicProcurement\ \\ 
\bottomrule
\end{tabular}
\end{table*}

Some DPO mistakes (such as \WrongDPOAdvice) may also cause a controller’s wrong assessment, which subsequently induces wrong organizational choices.

If a threat exploits even one vulnerability, it will determine a GDPR’s principles violation, exposing the controller to fines or penalties. Of course, this might depend on somebody tipping the supervisory authority or the SA coming to investigate its initiative or after a data breach.

\section{What does a DPO actually do?}\label{sec:DPOactions}

The boundary between the DPO's functions sometimes is fuzzy, especially in complex scenarios where more of them are involved. In Table \ref{tab:functionimpacted}, we summarized some examples, presenting the role of DPO in mitigating risks (namely: \emph{prevention}, \emph{detection}, \emph{response}, and \emph{investigation}) for each scenario from Table \ref{tab:scenarios}.

\begin{table*}
\centering
\caption{Some examples of activities that a DPO carries out to mitigate consequences of realized risks}\label{tab:functionimpacted}
\begin{tabular}{l@{~}p{0.65\textwidth}p{0.25\textwidth}}
\toprule
\textbf{ID} & \textbf{Illustrative mitigation by the DPO} & \textbf{Applicable Scenarios (or DPO's function)} \\
\midrule
\multicolumn{3}{l}{Prevention - Controller initiated} \\
\midrule
P1 & A group of joint controllers (two or more controllers who jointly determine the purposes and means of processing) asks their DPOs for advice on the technical and organizational aspects of periodic or new processing. For example, processing will build on an integrated territorial video surveillance system using OCR cameras. & \textit{Advisory function}, \textit{Cooperative function}, \textit{Organizational function} and \textit{Monitoring of compliance}\\
P2 & The DPO helps the controller to do a DPIA before starting new high-risk processing (e.g., one relating to a Covid-19 screening data acquisition and management system). & \textit{Organizational function} \\
P3 & The DPO supports the controller in choosing the configuration of a company's telecommunication equipment or a software tool that guarantees the protection of personal data by default. & \scriptsize \NoDataProtectionPrinciples, \UncheckedRemoteMonitoring \\
\midrule
\multicolumn{3}{l}{Prevention - DPO initiated} \\
\midrule
P4 & A DPO monitors the data protection laws changes and the indications of the bodies in charge, and after that, he prepares short information pills or notes for an SME’s staff. & \textit{Information and raising awareness function} \\
P5 & The DPO advises the controller that in issuing public tenders, it should expressly call for applicants that can “demonstrate” that their product or service fully complies with GDPR. & \scriptsize \WrongPublicProcurement \\
P6 & The DPO advises the controller to launch a census of all the PCs in the organization that have a Microsoft operating system version 7 or older. The DPO interacts with the ICT Department to develop an updated operating software plan. & \scriptsize \SoftwareEndofLife \\
P7 & The DPO of a National central bank illustrates to some banks' DPOs the controller's obligations. & \textit{Advisory function} \\
P8 & The DPO consult the competent SA about implementing a new data processing. & \textit{Cooperative function} \\
\midrule
\multicolumn{3}{l}{Investigate - DPO initiated} \\
\midrule
I1 & The DPO initiate an investigative activity to verify compliance with contract terms. & \scriptsize \SubcontractorViolatesPrivacy \\
I2 & The DPO immediately advise the controller about the existence of a violation of the contract terms. The controller, in turn, immediately formally warns the processor about this violation, ordering it to stop the infringement at once (e.g., by returning the data encryption key). & \scriptsize \SubcontractorViolatesPrivacy \\
\midrule
\multicolumn{3}{l}{Investigate - Data subject initiated} \\
\midrule
I3 & A DPO receives a piece of informal information and initiates an investigative activity (e.g., to verify the control procedure of the EU Digital COVID Certificate held by the staff). & \textit{Monitoring of compliance} and \textit{Handle queries or complaints} \\
\midrule
\multicolumn{3}{l}{Detection - DPO initiated} \\
\midrule
D1 & Examining the output of an audit, the DPO finds that the processing is not compliant with the GDPR principles or is unlawful. & \scriptsize \DPOAdviceNotSought, \WebsiteForcesChoices \\
D2 & The DPO conducts periodic audits of processing compliance with GDPR principles. & \scriptsize \IgnoredDPOAdvice, \DPOAdviceNotSought, \WebsiteForcesChoices \\
\midrule
\multicolumn{3}{l}{Response - Controller initiated} \\
\midrule
R1 & The DPO could refuse to sign the GDPR compliance of a new or modified processing. & \textit{Enforcement} \\
R2 & The DPO assist the controller in investigating if a personal data breach has occurred. & \scriptsize \DPOAdviceNotSought, \SubcontractorViolatesPrivacy \\
\midrule
\multicolumn{3}{l}{Response - DPO initiated} \\
\midrule
R3 & The DPO invite the controller to immediately remove irrelevant and excess personal data and apply the data minimization principle. & \scriptsize \AdminAsksForEverything, \WebsiteForcesChoices \\
R4 & The DPO interact with the controller's structure to follow up on the data subject's request. & \scriptsize \NeglectedSubjectRight, \SubjectRightRequest \\
R5 & The DPO advise the controller to immediately stop the processing by temporarily suspending the service to modify the software procedure. & \scriptsize \DPOAdviceNotSought, \WebsiteForcesChoices \\
R6 & After the controller follows up data subject request, the DPO reports it to the SA. & \scriptsize \NeglectedSubjectRight \\
R7 & The DPO notifies a personal data breach the SA, acting on behalf of the controller. & \scriptsize \DPOAdviceNotSought, \SubcontractorViolatesPrivacy \\
R9 & After the controller deletes irrelevant and excess personal data, the DPO notifies the data subject of complaint upholding. & \scriptsize \AdminAsksForEverything \\
R10 & The DPO communicate a data breach to the data subjects, acting on behalf of the controller. & \scriptsize \DPOAdviceNotSought, \SubcontractorViolatesPrivacy \\
\midrule
\multicolumn{3}{l}{Response - Supervisory Authority initiated}\\
\midrule
R8 & The DPO interact with the SA giving it the best collaboration in handling complaints lodged versus the controller and facilitating access to the documents and information. & \scriptsize \NeglectedSubjectRight \\
\midrule
\multicolumn{3}{l}{Response - Data subject initiated} \\
\midrule
R11 & The DPO checks if the complaint or request received from the data subject is well-founded. & \scriptsize \AdminAsksForEverything, \NeglectedSubjectRight, \SubjectRightRequest \\
R12 & The DPO responds to a data subject who applied for information about personal data processing, promptly providing all requested information. & \scriptsize \NeglectedSubjectRight, \SubjectRightRequest \\
\bottomrule
\end{tabular}
\end{table*}

In \DPOAdviceNotSought, the DPO's activities correspond to a detection (D1 and D2) and a response scenario (R2, R5, R7, and R10). The DPO primarily exercises the monitoring of compliance. However, in the response part, there is a combination of the advisory, organizational, and cooperative functions, performed secondarily, as well as the enforcement one.

The case \WebsiteForcesChoices\ corresponds to two different scenarios. The first is a detection scenario in which the DPO exercises the function monitoring of compliance (D1 and D2), while the second is a response one where the Enforcement function is applied (R5).

In the scenario \AdminAsksForEverything, the DPO carries out his or her activities in a response scenario. Some (R3 and R9) combine enforcement and handling queries or complaints functions. At the same time, another (R11) involves primarily the handling queries or complaints function and secondarily the monitoring of compliance one.

The importance of the DPO's cooperation function comes to light from the scenarios \NeglectedSubjectRight\ and \SubjectRightRequest. These cases are linked to two response scenarios. The first is primarily involved in the DPO's handling queries or complaints function, followed by the function monitoring of compliance (R11). The second relates to handling queries or complaints and monitoring compliance functions (R4 and R12). Moreover, related to \NeglectedSubjectRight, there is an additional response scenario in which the DPO’s cooperative function is the primary involved, which follows up handling queries or complaints function (R6 and R8). The cases \NoDataProtectionPrinciples\ and \UncheckedRemoteMonitoring\ highlight the importance of the DPO's organizational function (P3), primarily in a prevention scenario. The advisory function follows up as second in the same.

In \WrongPublicProcurement, there is an example of a prevention scenario in which the DPOs exercise their advisory function (P5). While in \SoftwareEndofLife, there is a prevention scenario where DPOs exercise a combination of their organizational and advisory functions (P6). The \SubcontractorViolatesPrivacy\ shows the DPO’s monitoring of compliance use in an investigative scenario (I1). Then, in a combination of response and investigative scenarios, the DPO exercises the monitoring of compliance function, followed by a mix of the advisory, organizational and cooperative functions performed secondarily (R2, R7, R10, and I2).

Regarding the DPO's advisory function, when DPOs are involved in new data processing, they can consult the SA if necessary (P8). In \WebsiteForcesChoices, when the DPO becomes aware of a process that is not entirely compliant with data protection policies (D2), he or she advises the controller, making recommendations for practically improving it (D1 and then R3 or R5).

Finally, the DPO should perform the investigative activity even if the controller does not involve him or her. For example, in \SubcontractorViolatesPrivacy, the DPO can independently carry out this activity (I1). If he or she finds a violation of data protection by design and default principles, the DPO acts accordingly (I2).

\section{What does a DPO need to know?}\label{sec:DPOskills}
GDPR does not specify the professional qualities required at a DPO, but only that the needed expert knowledge must be adequate for the data processing operations and their protection rank. The WP29 \cite{WP243rev01} suggests that it must be commensurate with the sensitivity, complexity, and amount of data the organization processes. Table \ref{tab:skills} summarizes the expertise and skills that a DPO should have. They consist of qualities, expertise in law and practices, ability, and educational qualification.

\begin{table*}
\centering
\caption{Expertise and skills of the DPO}\label{tab:skills}
\begin{minipage}{\textwidth}\footnotesize
This table describes the expertise and skills a DPO should have to carry out his or her tasks well.
\end{minipage}
\begin{tabular}{p{0.07\textwidth}p{0.20\textwidth}p{0.65\textwidth}}
\toprule
\textbf{Criteria} & 
\textbf{Description} & 
\textbf{Examples} \\
\midrule
Qualities &
DPOs must possess specific professional qualities & 
\emph{Supervisory authorities} provide continuous training courses reserved for DPOs (e.g., the T4Data international project and the SME Data project).

\emph{A controller} required in the call for DPO's appointment that the candidates must have: in-depth knowledge of the organizational structure, the information systems present, and the specific sector of activity of the controller, as well as being familiar with the data processing operations carried out by the latter.

\emph{EDPS} asserts that it is better to recruit the DPOs of EU institutions/bodies/agencies (EUI) within the EUI. These people usually ensure a better knowledge of the organization, structure, and functioning of the EUI itself.\\

Expert in law & 
DPOs must have expert knowledge of data protection law & 
\emph{The EDPS} asserts that the expert knowledge of data protection law is a prerequisite to the EUI's DPOs function.

\emph{A controller} required in the call for DPO's appointment that the candidates must know the legislation and practices on data protection both from a legal and IT point of view, including in-depth knowledge of the GDPR.\\

Expert in IT practice &
DPOs must have expert knowledge of IT, security, and organization &
According to \emph{EPDS} for EUI's DPOs, one of the professional qualities is knowledge of IT, including security aspects and organizational and communication skills.

\emph{The Network of Data Protection Officers} of the EU institutions and bodies recommends that the EUI's DPOs should have at least three years of relevant experience/maturity, to serve as DPO in a body where data protection is not related to the core business. Otherwise, this period grows to at least seven years. Similarly happens if the DPO will serve in an EU institution or which has an essential volume of processing operations. \\

Ability &
DPOs must have the ability to fulfill the tasks listed in GDPR, Article 39 &
\emph{A controller} explicitly required in the call for DPO's appointment that the candidates must have personal qualities, including integrity and high professional ethics.
According to \emph{EPDS} for EUI's DPOs, the DPOs' ability to fulfill their tasks should be referred to their personal qualities and knowledge and their position within the organization.\\

Educational qualification &
DPOs could have a variety of qualifications in law and computer science, security and privacy &
However, they cannot be uniquely determined. \emph{An Italian court ruling} asserts that holding ISO27001 certification cannot be a binding prerequisite in the selection procedure for a DPO's appointment.\\
\bottomrule
\end{tabular}
\end{table*}

In the absence of relevant bodies' specific guidance, from a legal perspective, it is challenging to define DPOs' selection criteria that can truly measure the adequacy of their level of knowledge. While technical and management skills are essential, there is no consensus on "specific" certifications that guarantees an adequate expert knowledge level of the DPO. As a result, it is complex to establish the absolute value of specific qualifications (e.g., university master's), professional certifications (e.g., UNI 11697: 2017 certification), and being an author of books, articles, papers, or research products. Courts reverted, as unfair requirements, several attempts to mandate this or that certification (e.g., BS-7799 or ISO 27001).

In our experience, a DPO can achieve a good knowledge of data protection practices by studying documents arranged by EDPB, EDPS, ENISA, and supervisory authorities of EU member states.

Another critical point is finding appropriate training for the professional profile of the DPO, from both a technical and a legal point of view: while knowledge of data protection law is a crucial requirement, a DPO may not have a law degree. A DPO may have a cybersecurity degree, but a European study \cite{Dragoni2021} 
found that European master of science (M.Sc.) programs in cybersecurity practically do not cover the knowledge units on component procurement. Knowing this unit is critical to guarantee compliance with the privacy-by-design principle because third-party components and contracts with IT providers are the norms for any administration because they rarely have in-house developers.

Some training support can also come from internal and external information sources. An example of internal ones may be information that a SOC (Security Operation Center) or the IT staff provides to DPO about events and incidents of security. The national CSIRT (Computer Security Incident Response Team) is an external source that provides pre-alerts, alerts, bulletins, and information regarding risks and incidents. For instance, in case of a data breach, a DPO should collaborate with the internal IT department (if available) and later refer to the national CSIRT.
As a further example, considering the interaction between a DPO and a SOC, the DPO will not have direct access to a security incident and event management (SIEM) system for an independent analysis of the inputs collected from the connected security devices and sensors \cite{bhatt2014}. Instead, the DPO will refer to summary reports prepared by security analysts. In the case of a data breach uncovering, direct contact with security analysts could help obtain more information about the incident and better determine its impact and extent. Unfortunately, in many places (e.g., small public administrations), the "IT security department" is just \emph{one} IT person who, among other duties, knows something about security. The DPO may end up being \emph{the} security expert.

The currently available technology can help DPOs make it easier to carry out their tasks. For example, using requirement analysis tools in software development or procurement could improve compliance with the data protection by design principle.

Additional support comes from research. Research findings can provide DPOs with insights into where to focus their efforts. According to this vision, the DPO advising task is “driven by research.” For example, Tang et al. \cite{tang2021} find that users have difficulties understanding the technical terms used in privacy policies because they misunderstand and misconstrue them. As a result, also the privacy policies themselves are misunderstood and misinterpreted. Considering the results of this and other similar studies helps the DPO understand the training gaps in the firm’s workforce.

\section{Conclusions}\label{sec:DPOrole}
The goal of compliance with the GDPR makes the DPO’s role dual. This data protection specialist is both the person who controls the processing activities in the organization and the person who acts as a wise advisor to the management. This tension could be problematic as the DPO needs information to carry out his or her duties. At the same time, the manager wants to have support in determining the purposes and means of processing personal data without giving the DPO too much information.

The duties assigned to the DPO role can quickly put this person in conflict within the organization for which she or he works. This case could happen especially in organizations with a negative attitude toward data protection.
For example, Hadar et al. \cite{hadar2018} reported a qualitative study where 17 developers out of 27 declared that the climate of the organization they worked for was averse to data protection. Developers reported that they must comply with organizational norms against data protection laws contradictory to the company's formally stated policies. In these circumstances, DPOs who carry out their duties will likely experience conflicts with the management. Casutt et al. \cite{casutt2020data} found a similar outcome. They showed that DPOs experienced an inherent conflict between complying with the law and realizing the organization's project whenever there is a gap between privacy requirements and those of the organization for which the DPO works.

A potential limitation of our research is that we based our scenarios on a detailed analysis of 90 specific cases, mostly of Italian origin, and court decisions may differ across EU Member States. Several factors mitigate this issue. At first, the relevant legislation (the GDPR) is a single regulation for all EU Member States and is directly applicable to them, regardless of the national legislation, and the \textit{European Data Protection Board} is tasked to facilitate the consistent application of data protection rules throughout the European Union and promote cooperation among the supervisory authorities of individual EU Member States. Second, we reviewed supervisory decisions of several countries \cite{Korff2019}, including over 1400 cases from \cite{EnforcementTracker}. So we are reasonably confident that our scenarios will stand the test of cross-border analysis.

While we do not have an engineering solution for the DPO's problems, at least being aware of the concrete problems is the first step toward a solution.

\ifCLASSOPTIONcompsoc

\section*{Acknowledgments}
\else
\section*{Acknowledgment}
\fi

The European Union has partly supported this work under the H2020 Leadership in enabling and industrial technologies (LEIT) program under grant agreement 830929 (CyberSec4Europe).

\subsection*{CRediT} statements
	\emph{Conceptualization:} FC, FM;
	\emph{Methodology:} FM, FC;
	\emph{Validation:} FM;
    \emph{Investigation:} FC;
    \emph{Data Curation:} FC;
    \emph{Writing - Original Draft:} FC;
    \emph{Writing - Review \& Editing:} FC, FM;
    \emph{Visualization:} FC;
    \emph{Supervision:} FM;
    \emph{Project administration:} FM;
    \emph{Funding acquisition:} FM.

\ifCLASSOPTIONcaptionsoff
  \newpage
\fi

\bibliographystyle{IEEEtran}

\input{dpo.bbl}

\clearpage

\twocolumn

\appendices

\section{Case Studies' References} \label{Appendix_Case_studies_references}\label{Supplemental_Material}

\subsection{Description of Supplementary Tables}
In distilling our reference scenarios, we followed the methodology suggested by Yin in \cite{yin2018case}. According to this vision, multiple (at least six) sources of evidence could exist for a case study building. 
Because better case studies rely on various sources usually used to build triangulation, we employed multiple sources of evidence (for example, court decisions, supervisory authorities' decisions, job advertisements, and newspaper articles). The distilled scenarios
mask the original persons or organization involved in the original case study to make it more general and less vulnerable to retraction based on the right to be forgotten.

Table \ref{tab:correlation} shows how each scenario from Table \ref{tab:scenarios} covers a DPO function as described in Table \ref{tab:function}. Table \ref{tab:consequences} links reference scenarios described in Table \ref{tab:scenarios} with an exemplification of possible impacts (consequences of the vulnerabilities) that are useful in better describing these scenarios.

Tables \ref{tab:case-studies-sources} and \ref{tab:case-studies-sources-2} lists the sources of case studies used to distill the scenario presented in Table \ref{tab:scenarios} (in Sections \ref{sec:DPOscenarios} and \ref{sec:TV}, respectively.).
Table \ref{tab:other-case-studies-sources} lists the sources of the other case study examples cited in the article that do not belong to the twelve scenarios.

The source of the example cited in Section \ref{sec:DPOskills} as a case of a court that reverted an attempt to mandate qualifications is the first one listed in Table \ref{tab:other-case-studies-sources}. For the years of professional experience in data protection that a DPO should have, one can refer to the second row of Table \ref{tab:other-case-studies-sources}. This document specifies that the head of the investigation expects the DPO to have at least three years of professional experience in data protection. Starting from the third row of the same reference in Table \ref{tab:other-case-studies-sources} are listed the related documents (such as the job descriptions and advertisements for hiring DPOs, and internal regulations describing the DPO's requirements) that we looked for in Section \ref{sec:DPOskills}.

In Table \ref{tab:OCR-camera-case-studies-sources} are listed the sources of evidence used to build the case related to an integrated video surveillance system by using an OCR camera (P1, Table \ref{tab:functionimpacted}).

Finally, we analyzed the GDPR's fines case types in the sites listed in \cite{EnforcementTracker}. We mapped the number of cases for each of them. Moreover, we related these types to our twelve scenarios (See Table\ref{tab:enforcement_tracker} for this classification).

\begin{table*}
\centering 
\caption{Scenarios' correlations to DPO functions}\label{tab:correlation}
\begin{minipage}{0.9\textwidth}
Table \ref{tab:correlation} shows how each scenario from Table \ref{tab:scenarios} covers a DPO function as described in Table \ref{tab:function}.
\end{minipage}

{\footnotesize
\begin{tabular}{p{0.25\textwidth}p{0.35\textwidth}p{0.35\textwidth}}
\toprule
\textbf{Short name} & 
\textbf{Primary Function(s)} & \textbf{Ancillary Function(s)} \\
\midrule
\WrongDPOAdvice & Advisory function & Organizational function \\
\IgnoredDPOAdvice & Organizational function, Advisory function & (lack of) Enforcement \\ 
\DPOAdviceNotSought &  Monitoring of compliance & Advisory function, Organizational function, Cooperative function, (lack of) Enforcement\\ 
\WebsiteForcesChoices & Monitoring of compliance & Enforcement, Advisory function \\

\AdminAsksForEverything & Handle queries or complaints (or) Monitoring of compliance & Monitoring of compliance (or) Handle queries or complaints, Enforcement, Information and raising awareness function \\ 
\NeglectedSubjectRight & Cooperative function, Handle queries or complaints & (lack of) Monitoring of compliance, (lack of) Organizational function \\ 
\SubjectRightRequest & Handle queries or complaints & Monitoring of compliance, Organizational function \\ 
\NoDataProtectionPrinciples & Organizational function & Advisory function \\ 
\UncheckedRemoteMonitoring & Organizational function & Advisory function \\ 
\WrongPublicProcurement & Advisory function & Information and raising awareness function\\ 
\SoftwareEndofLife & Advisory function & Organizational function  \\
\SubcontractorViolatesPrivacy & Investigation, Monitoring of compliance & Advisory function, Organizational function, Enforcement, Cooperative function \\
\bottomrule
\end{tabular}}
\end{table*}

\begin{table*}
\centering 
\caption{Scenarios' possibles consequences}\label{tab:consequences}
\begin{minipage}{0.9\textwidth}
Table \ref{tab:consequences} links reference scenarios described in Table \ref{tab:scenarios} with an exemplification of possible       impacts (consequences of the vulnerabilities) that are useful in better describing these scenarios.
\end{minipage}

{\footnotesize
\begin{tabular}{p{0.25\textwidth}p{0.65\textwidth}}
\toprule
\textbf{Short name} & 
\textbf{Examples of consequences}\\
\midrule
\WrongDPOAdvice & The controller evaluation of the risk of processing on rights and freedom of individuals could be wrong.\\
\hline
\IgnoredDPOAdvice & The processing could be risky for the rights and freedoms of natural persons. \\ 
\hline
\DPOAdviceNotSought & Anyone could pretend to be another data subject and access his or her personal data. \\ 
\hline
\WebsiteForcesChoices & It is lost the legal basis of the processing.\\
\hline
\AdminAsksForEverything & Excess and irrelevant personal data are disseminated over the Internet by the controller, making them accessible to an indefinite number of unauthorized subjects. \\ 
\hline
\NeglectedSubjectRight & The infringement of the data subjects' rights provisions expose the controller to administrative fines up to 20000000 EUR, or up to 4\% of the total worldwide annual turnover. \\ 
\hline
\SubjectRightRequest & The controller may not be able to respond to the data subject's request, exposing it to complaints and administrative fines. \\ 
\hline
\NoDataProtectionPrinciples & A personal data breach may happen because of unauthorized disclosure of personal data. \\ 
\hline
\UncheckedRemoteMonitoring & The technicians can connect to a workstation independently without notifying their access request user and acquiring its preventive consent. Moreover, they can connect to a workstation, stealthy monitoring all activities the user logged into the computer is doing. \\
\hline 
\WrongPublicProcurement & A contractor that offers products or services, not GDPR-compliant can win a bid, and the call for tender may be subject to litigation. \\ 
\hline
\SoftwareEndofLife & It is impossible to guarantee adequate security and functionality of the equipment. \\ 
\hline
\SubcontractorViolatesPrivacy & The tender may be subject to litigation. \\
\bottomrule
\end{tabular}}
\end{table*}

\vfill$\,$
\newpage

\begin{table*}
\centering
\caption{Document' sources of DPO’s functions description section} \label{tab:DPO-functions-description-sources}

\begin{tabular}{p{0.2\textwidth}p{0.7\textwidth}}
\toprule
Document source & 
Document description\\
\midrule \scriptsize
European Data Protection Supervisor & “Position paper on the role of Data Protection Officers in ensuring effective compliance with Regulation (EC) 45/2001,” Brussels, pp. 1–11, 2005. \\
Article 29 Data Protection Working Party & “Guidelines on Data Protection Impact Assessment (DPIA) and determining whether processing is “likely to result in a high risk” for the purposes of Regulation 2016/679,” pp. 1–22, 2017. \\

Il Manifesto (newspaper) & “Marche, il buco dello screening: dati accessibili a chiunque,” 2021. [Online]. Available: \url{https://ilmanifesto.it/marche-il-buco-dello-screening-dati-accessibili-a-chiunque/} [English translation: Marche, the screening hole: data accessible to anyone].\\

Garante per la protezione dei dati personali & “Ordinanza ingiunzione nei confronti di Azienda sanitaria unica regionale Marche - 13 gennaio 2022 [9744496],” Roma, 2022. [Online]. Available: \url{https://www.garanteprivacy.it/web/guest/home/docweb/-/docweb-display/docweb/9744496} [English translation: Injunction order against single regional health company Marche - Jan. 13, 2022].\\

European Commission & “EU Digital COVID Certificate.” [Online]. Available: \url{https://ec.europa.eu/info/live-work-travel-eu/coronavirus-response/safe-covid-19-vaccines-europeans/eu-digital-covid-certificate_en} \\

European Parliament and of the Council & “Regulation (EU) 2021/953 of the European Parliament and of the Council of 14 June 2021 on a framework for the issuance, verification and acceptance of interoperable COVID-19 vaccination, test and recovery certificates (EU Digital COVID Certificate) to facilitate free movement during the COVID-19 pandemic,” pp. 3–25, jun 2021. \\

Repubblica Italiana & “DECRETO-LEGGE 22 aprile 2021, n. 52. Misure urgenti per la graduale ripresa delle attivit`a economiche e sociali nel rispetto delle esigenze di contenimento della diffusione dell’epidemia da COVID-19.” pp. 3–12, apr 2021. [English translation: DECREE-LAW No. 52 of April 22, 2021. Urgent measures for the gradual resumption of economic and social activities in compliance with the needs to contain the spread of the COVID-19 epidemic.].\\

Azienda sanitaria unica regionale delle Marche) & “Regolamento organizzativo aziendale privacy dell’Azienda sanitaria unica regionale delle Marche (Determina DG n. 349 del 30/05/2018,” Ancona, 2018. [English translation: Corporate organizational privacy regulations of the Single Regional Health Authority of the Marche Region (DG Determination No. 349 of 30/05/2018)]. \\

Banca d’Italia & “Relazione del Responsabile della Protezione dei Dati - anno 2020,” Roma, Tech. Rep., 2020. [English translation: Report of the Data Protection Officer].\\

\bottomrule
\end{tabular}
\end{table*}

\vfill$\,$
\newpage

\begin{table*}
\centering
\caption{Document' sources of case studies described in Table \ref{tab:scenarios}, Section \ref{sec:DPOscenarios}} \label{tab:case-studies-sources}

\begin{tabular}{p{0.10\textwidth}p{0.85\textwidth}}
\toprule
Case study & 
Document' sources\\
\midrule

\scriptsize \WrongDPOAdvice & \scriptsize Provvedimento del Garante per la protezione dei dati personali n. 118 del 2 luglio 2020 [9440025]. Available: \url{https://www.garanteprivacy.it/home/docweb/-/docweb-display/docweb/9440025} \\
 & \scriptsize Provvedimento del Garante per la protezione dei dati personali n. 317 del 16 settembre 2021 [9703988]. Available: \url{https://www.garanteprivacy.it/web/guest/home/docweb/-/docweb-display/docweb/9703988} \\
\hline

\scriptsize \IgnoredDPOAdvice & \scriptsize Provvedimento del Garante per la protezione dei dati personali n. 207 del 25 maggio 2021 [9590466]. Available: \url{https://www.garanteprivacy.it/web/guest/home/docweb/-/docweb-display/docweb/9590466} \\
 & \scriptsize Risposta a un quesito sull’identificazione degli intestatari del Green Pass del Garante per la protezione dei dati personali [9688875]. Available: \url{https://www.garanteprivacy.it/web/guest/home/docweb/-/docweb-display/docweb/9688875} \\
 & \scriptsize Parere del Garante per la protezione dei dati personali sul DPCM di attuazione della piattaforma nazionale DGC per l'emissione, il rilascio e la verifica del Green Pass - 9 giugno 2021 [9668064]. Available: \url{https://www.garanteprivacy.it/web/guest/home/docweb/-/docweb-display/docweb/9668064} \\
 & \scriptsize Provvedimento del Garante per la protezione dei dati personali n. 170 del 7 aprile 2022 [9773687]. Available: \url{https://www.garanteprivacy.it/web/guest/home/docweb/-/docweb-display/docweb/9773687} \\
 & \scriptsize Provvedimento del Garante per la protezione dei dati personali n. 273 del 22 luglio 2021 [9683814]. Available: \url{https://www.garanteprivacy.it/web/guest/home/docweb/-/docweb-display/docweb/9683814} \\
 & \scriptsize Provvedimento del Garante per la protezione dei dati personali n. 430 del 13 dicembre 2021 [9727220]. Available: \url{https://www.garanteprivacy.it/web/guest/home/docweb/-/docweb-display/docweb/9727220} \\
 & \scriptsize Green Pass: le ultime indicazioni del Garante Privacy, ma rimangono aspetti da chiarire \url{https://www.cybersecurity360.it/legal/privacy-dati-personali/green-pass-le-ultime-indicazioni-del-garante-privacy-ma-rimangono-aspetti-da-chiarire/} \\
 & \scriptsize Green pass, controlli durante l’orario di servizio o al termine delle lezioni. Lo Snals dice stop: “Manca il rispetto della privacy” \url{https://www.orizzontescuola.it/green-pass-controlli-durante-lorario-di-servizio-o-al-termine-delle-lezioni-lo-snals-dice-stop-manca-il-rispetto-della-privacy/} \\ 
 & \scriptsize Segnalazione del Garante per la protezione dei dati personali al Parlamento e al Governo sul Disegno di legge di conversione del decreto-legge n. 127 del 2021 (AS 2394), in relazione alla possibilità di consegna, da parte dei lavoratori dei settori pubblico e privato, di copia della certificazione verde, al datore di lavoro, con la conseguente esenzione, dai controlli, per tutta la durata della validità del certificato [9717878]. Available: \url{https://www.garanteprivacy.it/web/guest/home/docweb/-/docweb-display/docweb/9717878}\\
\hline

\scriptsize \DPOAdviceNotSought & \scriptsize Scambio sacche per trapianto midollo a San Martino di Genova [English translation: "Marrow transplant bag exchange at San Martino in Genoa, Italy"]. Available online: \url{https://genova.repubblica.it/cronaca/2018/10/17/news/scambio_sacche_per_trapianto_midollo_a_san_martino_di_genova-209207054/} \\
 & \scriptsize Vimercate, donna muore in ospedale dopo trasfusione: sacche di sangue scambiate per omonimia [English translation: "Vimercate, woman dies in hospital after transfusion: blood bags mistaken for homonymy"]. Available online:\url{https://www.ilmessaggero.it/italia/monza_ospedale_sacche_sangue_trasfusione_errore_donna_morta-4737967.html?refresh_ce} \\
\hline

\scriptsize \WebsiteForcesChoices & \scriptsize Provvedimento del Garante per la protezione dei dati personali n. 119 del 27 gennaio 2021 [9711630]. Available: \url{https://www.garanteprivacy.it/web/guest/home/docweb/-/docweb-display/docweb/9711630} \\
 & \scriptsize Provvedimento del Garante per la protezione dei dati personali n. 317 del 16 settembre 2021 [9703988]. Available: \url{https://www.garanteprivacy.it/web/guest/home/docweb/-/docweb-display/docweb/9703988} \\
 & \scriptsize Provvedimento del Garante per la protezione dei dati personali n. 7 del 15 gennaio 2020 [9256486]. Available: \url{https://www.garanteprivacy.it/web/guest/home/docweb/-/docweb-display/docweb/9256486} \\
 & \scriptsize Provvedimento del Garante per la protezione dei dati personali n. 195 del 26 maggio 2022 [9788986]. Available: \url{https://www.garanteprivacy.it/web/guest/home/docweb/-/docweb-display/docweb/9788986} \\
 \hline
 
\scriptsize \AdminAsksForEverything & \scriptsize Provvedimento del Garante per la protezione dei dati personali n. 173 del 1 ottobre 2020 [9483375],” Roma, 2020. [Online]. Available: \url{https://www.garanteprivacy.it/web/guest/home/docweb/-/docweb-display/docweb/9483375} \\
 & \scriptsize Provvedimento del Garante per la protezione dei dati personali n. 272 del 17 dicembre 2020 [9557593],” Roma, 2020. [Online]. Available: \url{https://www.garanteprivacy.it/web/guest/home/docweb/-/docweb-display/docweb/9557593} \\
 & \scriptsize Provvedimento del Garante per la protezione dei dati personali n. 204 del 26 maggio 2022 [9780409]. Available: \url{https://www.garanteprivacy.it/web/guest/home/docweb/-/docweb-display/docweb/9780409} \\
 & \scriptsize Dati personali pubblicati online: interviene il Garante. Come difenderci dagli abusi. [English translation: Personal data published online: the Guarantor intervenes. How to defend against abuse]. Available: \url{https://www.repubblica.it/economia/diritti-e-consumi/diritti-consumatori/2022/06/19/news/dati_personali_pubblicati_online_interviene_il_garante_come_difenderci_dagli_abusi-354299516/} \\
\hline

\scriptsize \NeglectedSubjectRight & \scriptsize Provvedimento del Garante per la protezione dei dati personali n. 40 del 26 febbraio 2020 [9365135],” Roma, 2020. [Online]. 
Available:  \url{https://garanteprivacy.it/web/guest/home/docweb/-/docwebdisplay/docweb/9365135} \\
 & \scriptsize Provvedimento del Garante per la protezione dei dati personali n. 174 del 12 maggio 2022 [9781242]. Available: \url{https://www.garanteprivacy.it/web/guest/home/docweb/-/docweb-display/docweb/9781242} \\
 & \scriptsize Provvedimento del Garante per la protezione dei dati personali n. 54 dell'11 febbraio 2021 [9556625]. Available: \url{https://www.garanteprivacy.it/web/guest/home/docweb/-/docweb-display/docweb/9556625} \\
 & \scriptsize Provvedimento del Garante per la protezione dei dati personali n. 7 del 15 gennaio 2020 [9256486]. Available: \url{https://www.garanteprivacy.it/web/guest/home/docweb/-/docweb-display/docweb/9256486} \\
\hline
 
\scriptsize \SubjectRightRequest & \scriptsize Provvedimento del Garante per la protezione dei dati personali n. 280 del 17 dicembre 2020, Roma, 2020. Available: \url{https://www.garanteprivacy.it/web/guest/home/docweb/-/docweb-display/docweb/9524175} \\
 & \scriptsize Provvedimento del Garante per la protezione dei dati personali n. 174 del 12 maggio 2022 [9781242]. Available: \url{https://www.garanteprivacy.it/web/guest/home/docweb/-/docweb-display/docweb/9781242} \\
 & \scriptsize Provvedimento del Garante per la protezione dei dati personali n. 54 dell'11 febbraio 2021 [9556625]. Available: \url{https://www.garanteprivacy.it/web/guest/home/docweb/-/docweb-display/docweb/9556625} \\
 \hline

\bottomrule
\end{tabular}
\end{table*}

\vfill$\,$
\newpage

\begin{table*}
\centering
\caption{Document' sources of case studies described in Table \ref{tab:scenarios}, Section \ref{sec:TV} }\label{tab:case-studies-sources-2}

\begin{tabular}{p{0.10\textwidth}p{0.85\textwidth}}
\toprule
Case study & 
Document' sources\\
\midrule

\scriptsize \NoDataProtectionPrinciples & \scriptsize Provvedimento del Garante per la protezione dei dati personali n. 285 del 22 luglio 2021 [9685994]. Available: \url{https://www.garanteprivacy.it/web/guest/home/docweb/-/docweb-display/docweb/9685994} \\
 & \scriptsize Provvedimento del Garante per la protezione dei dati personali n. 9 del 13 gennaio 2022 [9744496]. Available: \url{https://www.garanteprivacy.it/web/guest/home/docweb/-/docweb-display/docweb/9744496} \\
 & \scriptsize (Il Manifesto), "Marche, il buco dello screening: dati accessibili a chiunque", 2021. [English translation: Marche, the screening hole: data accessible to anyone]. Available: \url{https://ilmanifesto.it/marche-il-buco-dello-screening-dati-accessibili-a-chiunque} \\
 & \scriptsize Garante privacy, multa da un milione ad Atac e Comune di Roma: "Dati degli automobilisti non tutelati". [English translation: Privacy Guarantor, one million fine to Atac and Rome Municipality: 'Motorists' data not protected']. Available: \url{https://roma.repubblica.it/cronaca/2021/09/10/news/garante_privacy_multa_atac_e_comune-317239933/} \\
\hline

\scriptsize \UncheckedRemoteMonitoring & \scriptsize Provvedimento del Garante per la protezione dei dati personali n. 190 del 13 maggio 2021 [9669974]. Available: \url{https://www.garanteprivacy.it/web/guest/home/docweb/-/docweb-display/docweb/9669974} \\
 & \scriptsize Provvedimento del Garante per la protezione dei dati personali n. 137 del 15 aprile 2021 [9670738]. Available: \url{https://www.garanteprivacy.it/web/guest/home/docweb/-/docweb-display/docweb/9670738} \\
 & \scriptsize CEDU, Judgment 5 September 2017, CASE OF BĂRBULESCU v. ROMANIA (Application no. 61496/08). Available: \url{https://hudoc.echr.coe.int/eng#{%22itemid%22:[%22001-177082%22]}} \\
\hline

\scriptsize \WrongPublicProcurement & \scriptsize Provvedimento del Garante per la protezione dei dati personali n. 16 del 30 gennaio 2020 [9283857]. Available: \url{https://www.garanteprivacy.it/web/guest/home/docweb/-/docweb-display/docweb/9283857} \\
\hline

\scriptsize \SoftwareEndofLife & \scriptsize Provvedimento del Garante per la protezione dei dati personali n. 83 del 4 aprile 2019 [9101974]. Available: \url{https://www.garanteprivacy.it/web/guest/home/docweb/-/docweb-display/docweb/9101974} \\
& \scriptsize Provvedimento del Garante per la protezione dei dati personali n. 548 del 21 dicembre 2017 [7400401]. Available: \url{https://www.garanteprivacy.it/web/guest/home/docweb/-/docweb-display/docweb/7400401} \\
 & \scriptsize Software dispositivi medici è obsoleto. [English translation: Medical device software is obsolete]. Available: \url{https://www.ansa.it/sito/notizie/tecnologia/software_app/2015/06/22/software-dispositivi-medici-e-obsoleto_98a87cdc-accd-4f8d-b500-5b1ef7cce47d.html} \\
 & \scriptsize Cybersecurity,89\% strutture sanità Italia ha software datati. [English translation: Cybersecurity,89\% healthcare facilities Italy has outdated software]. Available: \url{https://www.ansa.it/sito/notizie/tecnologia/hitech/2021/12/07/cybersecurity89-strutture-sanita-italia-ha-software-datati_b5eda4c6-73a6-47f4-a007-b8930aba084b.html} \\

\hline

\scriptsize \SubcontractorViolatesPrivacy & \scriptsize Provvedimento del Garante per la protezione dei dati personali n. 16 del 30 gennaio 2020 [9283857]. Available: \url{https://www.garanteprivacy.it/web/guest/home/docweb/-/docweb-display/docweb/9283857} \\
 & \scriptsize Provvedimento del Garante per la protezione dei dati personali n. 43 del 10 febbraio 2022 [9751498]. Available: \url{https://www.garanteprivacy.it/web/guest/home/docweb/-/docweb-display/docweb/9751498} \\
 & \scriptsize Provvedimento del Garante per la protezione dei dati personali n. 44 del 10 febbraio 2022 [9754332]. Available: \url{https://www.garanteprivacy.it/web/guest/home/docweb/-/docweb-display/docweb/9754332} \\
 
\bottomrule
\end{tabular}
\end{table*}

\begin{table*}
\centering
\caption{Document' sources of case studies mentioned in the paper extending the scenarios described in Table \ref{tab:scenarios}} \label{tab:other-case-studies-sources}

\begin{tabular}{p{0.10\textwidth}p{0.85\textwidth}}
\toprule
References & 
Document' sources\\
\midrule

\scriptsize Side Channels & \scriptsize (Youtvrs), “Tolentino, il video choc dell’incidente alla rotonda in viale Buozzi,” 2019. [Online]. Available: \url{https://www.youtvrs.it/tolentino-il-video-choc-dellincidente-alla-rotonda-in-viale-buozzi/} [English translation: "Tolentino, shocking video of the accident at the traffic circle on Buozzi Avenue]. \\
& \scriptsize Provvedimento del Garante per la protezione dei dati personali n. 163 del 28 aprile 2022 [9777996]. Available: \url{https://www.garanteprivacy.it/web/guest/home/docweb/-/docweb-display/docweb/9777996} \\
\hline

\scriptsize Section \ref{sec:impactsanalysis}, Table \ref{tab:vulnerabilitiesimpact} & \scriptsize Provvedimento del Garante per la protezione dei dati personali n. 419 del 2 dicembre 2021 [9733053]. Available: \url{https://www.garanteprivacy.it/web/guest/home/docweb/-/docweb-display/docweb/9733053} \\
 & \scriptsize Provvedimento del Garante per la protezione dei dati personali n. 277 del 22 luglio 2021 [9693442]. Available: \url{https://www.garanteprivacy.it/web/guest/home/docweb/-/docweb-display/docweb/9693442} \\
 & \scriptsize Provvedimento del Garante per la protezione dei dati personali n. 83 del 4 aprile 2019 [9101974]. Available: \url{https://www.garanteprivacy.it/web/guest/home/docweb/-/docweb-display/docweb/9101974} \\
 & \scriptsize Attacco hacker: furto di dati personali ai clienti EasyCoop. [English translation: Hacker attack: theft of personal data from EasyCoop customers]. Available: \url{https://www.repubblica.it/economia/diritti-e-consumi/diritti-consumatori/2022/06/09/news/attacco_hacker_furto_di_dati_personali_ai_clienti_easycoop-353134656/} \\
 & \scriptsize Germania, attacco hacker a ospedale ha provocato morte donna. [English translation: Germany, hacker attack on hospital resulted in woman's death]. Available: \url{https://www.ansa.it/sito/notizie/tecnologia/internet_social/2020/09/18/germania-attacco-hacker-a-ospedale-ha-provocato-morte-donna_663eee64-a732-4fb5-84b2-863335ed9b22.html} \\
 & \scriptsize Attacco hacker, dopo la Bicocca allarme per i servizi sanitari. [English translation: Hacker attack, after Bicocca alert for health services]. Available: \url{https://milano.corriere.it/notizie/cronaca/17_maggio_13/attacco-hacker-la-bicocca-allarme-servizi-sanitari-349880d4-37d1-11e7-ad4d-8609abc1aa8e.shtml} \\
 & \scriptsize Attacco hacker Regione Lazio, ripartono solo le prenotazioni dei vaccini. Ecco l’elenco di tutti i servizi non ancora accessibili. [English translation: Lazio Region hacker attack, only vaccine reservations restart. Here's a list of all services not yet accessible]. Available: \url{https://www.ilfattoquotidiano.it/2021/08/06/attacco-hacker-regione-lazio-ripartono-solo-le-prenotazioni-dei-vaccini-ecco-lelenco-di-tutti-i-servizi-non-ancora-accessibili/6284498/} \\
 & \scriptsize Regione Lazio, a un mese dall’attacco hacker ancora disagi ai sistemi informatici. Fatture, tamponi e green pass: i servizi rallentati. [English translation: Lazio Region, one month after hacker attack still disrupting IT systems. Bills, buffers and green passes: services slowed down]. Available:\url{https://www.ilfattoquotidiano.it/2021/09/03/regione-lazio-a-un-mese-dallattacco-hacker-ancora-disagi-ai-sistemi-informatici-fatture-tamponi-e-green-pass-i-servizi-rallentati/6308625/} \\
 & \scriptsize Il Mise ha nascosto un furto di dati dai suoi sistemi per mesi. [English translation: Mise hid data theft from its systems for months]. Available: \url{https://www.wired.it/internet/web/2021/03/16/furto-dati-mise-hackerato-ministero/} \\
 & \scriptsize Uber, rubati i dati di 57 milioni di utenti e autisti. E l'azienda ha insabbiato tutto. [English translation: Uber, data of 57 million users and drivers stolen. And the company covered it up]. Available: \url{https://www.wired.it/attualita/tech/2017/11/22/cyberattacco-uber-riscatto/} \\
 & \scriptsize Dati di clienti e autisti non protetti: multe contro Uber per le falle nella privacy. [English translation: Unprotected customer and driver data: fines against Uber for privacy flaws]. Available: \url{https://www.wired.it/internet/regole/2018/11/27/dati-uber-attacco-privacy-multa/} \\
 & \scriptsize Il Garante della privacy ha multato Uber per 4 milioni di euro. [English translation: Privacy watchdog fined Uber 4 million euros]. Available: \url{https://www.wired.it/article/il-garante-della-privacy-ha-multato-uber-per-4-milioni-di-euro/} \\
 & \scriptsize Provvedimento del Garante per la protezione dei dati personali n. 101 del 24 marzo 2022 [9771142]. Available: \url{https://www.garanteprivacy.it/web/guest/home/docweb/-/docweb-display/docweb/9771142} \\
 & \scriptsize Provvedimento del Garante per la protezione dei dati personali n. 127 del 7 aprile 2022 [9771545]. Available: \url{https://www.garanteprivacy.it/web/guest/home/docweb/-/docweb-display/docweb/9771545} \\
 & \scriptsize Provvedimento del Garante per la protezione dei dati personali n. 49 del 10 aprile 2022 [9756869]. Available: \url{https://www.garanteprivacy.it/web/guest/home/docweb/-/docweb-display/docweb/9756869} \\
\hline

\scriptsize Section \ref{sec:DPOskills} & \scriptsize Sentenza del Tribunale Amministrativo del Friuli-Venezia Giulia, la n.287 del 13 settembre 2018, Trieste, 2018. [English translation: "Judgment of the Administrative Court of Friuli-Venezia Giulia, No. 287 of September 13, 2018"]. Available: \url{https://www.giustizia-amministrativa.it/portale/pages/istituzionale/ucm?id=5LLMWH2MBE2JVPC536FUMJHNYU&q} \\
 & \scriptsize Decision of the National Commission sitting in restricted formation on the outcome of investigation no. [...] carried out at public establishment A Deliberation no. 38FR/2021 of October 15, 2021 [original text in French: Décision de la Commission nationale siégeant en formation restreinte sur l’issue de l’enquête n° […] menée auprès de l’établissement public A Délibération n° 38FR/2021 du 15 octobre 2021].  Available: \url{https://cnpd.public.lu/content/dam/cnpd/fr/decisions-fr/2021/Decision-38FR-2021-sous-forme-anonymisee.pdf} \\
 & \scriptsize Regolamento protezione dati personali della Provincia di Fermo \url{https://www.provincia.fermo.it/public/2019/12/13/1\_regolamento-per-il-trattamento-dei-dati-personali.pdf} \\
 & \scriptsize Regolamento comunale per l'attuazione del Regolamento UE 2016/679 relativo alla protezione delle persone fisiche con riguardo al trattamento dei dati personali del Comune di Tolentino \url{https://www.comune.tolentino.mc.it/wp-content/blogs.dir/2/files/regolamento-trattam.-dati-personalii.pdf} \\
 & \scriptsize Delibera del Consiglio Comunale di Tolentino n. 12 del 5 marzo 2020 \url{https://www.halleyweb.com/c043053/zf/index.php/atti-amministrativi/delibere/dettaglio/atto/G1XpVNET6UT0-A} \\
 & \scriptsize Università degli Studi di Macerata, Decreto rettorale n. 137 prot. n. 11585 del 17 aprile 2019 \url{https://www.unimc.it/it/ateneo/direttore-generale/ufficio-legale/dr-137-2019.pdf} \\
 & \scriptsize Direttiva del Ministero dello Sviluppo Economico prot. n. U0002663 del 28 gennaio 2020 \url{https://www.mise.gov.it/images/stories/trasparenza/2020/Direttiva-in-materia-di-dati-personali-2020.pdf} \\
 & \scriptsize Determina del Direttore Generale ASUR n. 349 del 30 maggio 2018 (Regolamento organizzativo aziendale privacy dell'Azienda Sanitaria Unica Regionale delle Marche) - Allegato 1 \url{https://serviziweb.asur.marche.it/ALBI/ASUR2018/allegati/349DG - allegato 1.pdf} \\
 & \scriptsize Determina del Direttore Generale ASUR n. 349 del 30 maggio 2018 - Allegato 2 \url{https://serviziweb.asur.marche.it/ALBI/ASUR2018/allegati/349DG - allegato 2.pdf} \\
 & \scriptsize Determina del Direttore Generale ASUR n. 349 del 30 maggio 2018 \url{https://serviziweb.asur.marche.it/ALBI/ASUR2018/allegati/349DG(6).pdf} \\
 & \scriptsize Sentenza del Tribunale Amministrativo Regionale per il Friuli Venezia Giulia n. 00287 del 13 settembre 2018 \url{https://www.giustizia-amministrativa.it/portale/pages/istituzionale/ucm?id=5LLMWH2MBE2JVPC536FUMJHNYU&q} \\
 & \scriptsize Università degli Studi di Camerino, Bando di selezione interna per il conferimento dell'incarico di responsabile della protezione dei dati (RPD) per l'Ateneo ai sensi dell'art. 37 del Regolamento UE 2016/679 (GDPR) \url{https://www.unicam.it/sites/default/files/bandi/2018/04/Bando selezione RPD.pdf} \\
 & \scriptsize Determina del Direttore Generale ASUR n. 256 del 26 aprile 2018 \url{https://serviziweb.asur.marche.it/ALBI/ASUR2018/allegati/256DG(5).pdf} \\
 & \scriptsize Determina del Direttore Generale ASUR n. 256 del 26 aprile 2018 - Allegato 1 \url{https://serviziweb.asur.marche.it/ALBI/ASUR2018/allegati/256DG - all(1).pdf} \\

\bottomrule
\end{tabular}
\end{table*}

\begin{table*}
\centering
\caption{Document' sources of case study relate to an integrated video-surveillance system, referred in Table \ref{tab:functionimpacted}} \label{tab:OCR-camera-case-studies-sources}

\begin{tabular}{p{0.15\textwidth}p{0.80\textwidth}}
\toprule
Source type & 
Document' sources\\
\midrule

\scriptsize Online newspaper & \scriptsize Patti per videosorveglianza, la firma di 21 Comuni. [English translation: Covenants for video surveillance, the signing of 21 municipalities.]. Available: \url{https://www.cronachemaceratesi.it/2021/12/09/patti-per-videosorveglianza-la-firma-di-21-comuni/1591504/} \\
\scriptsize Online newspaper & \scriptsize Videosorveglianza, Camerano aderisce al protocollo di intesa di Macerata. [English translation: Video surveillance, Camerano joins Macerata's memorandum of understanding]. Available: \url{https://www.cronacheancona.it/2019/12/06/videosorveglianza-camerano-aderisce-al-protocollo-di-intesa-di-macerata/204561/} \\
\scriptsize Online newspaper & \scriptsize  Sistemi di videosorveglianza in rete: i Comuni riuniti a Macerata per rafforzare la sicurezza. [English translation: Networked video surveillance systems: municipalities gather in Macerata to strengthen security]. Available: \url{https://www.cronacheancona.it/2019/06/04/sistemi-di-videosorveglianza-in-rete-i-comuni-riuniti-a-macerata-per-rafforzare-la-sicurezza/170249/} \\
\scriptsize Online newspaper & \scriptsize  Telecamere e lettura delle targhe, al vaglio l'integrazione dei sistemi di videosorveglianza. [English translation: Cameras and license plate reading under consideration for integration of video surveillance systems]. Available: \url{https://www.cronacheancona.it/2019/05/30/telecamere-e-lettura-delle-targhe-al-vaglio-lintegrazione-dei-sistemi-di-videosorveglianza/169125/} \\
\scriptsize Online newspaper & \scriptsize  Verso un sistema integrato di videosorveglianza urbana: tavola rotonda a Macerata per mettere a fuoco obiettivi e strategie comuni. [English translation: Toward an integrated urban video surveillance system: roundtable in Macerata to focus on common goals and strategies]. Available: \url{https://www.tmnotizie.com/verso-un-sistema-integrato-di-videosorveglianza-urbana-tavola-rotonda-a-macerata-per-mettere-a-fuoco-obiettivi-e-strategie-comuni/} \\
\scriptsize Online newspaper & \scriptsize  Videosorveglianza, Grottammare dice sì al Comune di Macerata per l’uso condiviso dei dati. [English translation: Video surveillance, Grottammare says yes to Macerata municipality for shared use of data]. Available: \url{http://www.ilgraffio.online/2020/02/06/videosorveglianza-grottammare-dice-si-al-comune-macerata-luso-condiviso-dei-dati/} \\
\scriptsize Online newspaper & \scriptsize  Telecamere in rete per tanti Comuni di Ancona, Macerata e Fermo. Obiettivo sicurezza. [English translation: Networked cameras for many municipalities in Ancona, Macerata and Fermo. Security goal]. Available: \url{https://www.centropagina.it/osimo/telecamere-in-rete-per-tanti-comuni-di-ancona-macerata-e-fermo-obiettivo-sicurezza/} \\
\scriptsize Online newspaper & \scriptsize  La videosorveglianza va ko: i topi masticano la fibra ottica. [English translation: Video surveillance goes down: rats chew through fiber optics]. Available: \url{https://www.cronachemaceratesi.it/2021/11/06/la-videosorveglianza-va-ko-i-topi-masticano-la-fibra-ottica/1580995/} \\
\scriptsize Memorandum of understanding & \scriptsize  Protocollo di intesa per la gestione del sistema di videosorveglianza del Comune di Macerata. [English translation: Memorandum of understanding for the management of the video surveillance system of the Municipality of Macerata]. Available: \url{http://www.prefettura.it/macerata/download.php?coming=Y29udGVudXRpL1Byb3RvY29sbG9fZGlfaW50ZXNhX3Blcl9sYV9nZXN0aW9uZV9kZWxfc2lzdGVtYV9kaV92aWRlb3NvcnZlZ2xpYW56YV9kZWxfY29tdW5lX2RpX21hY2VyYXRhLTk1MzU4MDguaHRt&f=Spages&file=L0ZJTEVTL0FsbGVnYXRpUGFnLzExOTMvUFJPVE9DT0xMT1NDQ05UVE1BQ0VSQVRBMjNsdWcyMDIwLnBkZg==&id_sito=1193&s=download.php} and \url{http://www.prefettura.it/macerata/download.php?coming=Y29udGVudXRpL1Byb3RvY29sbG9fZGlfaW50ZXNhX3Blcl9sYV9nZXN0aW9uZV9kZWxfc2lzdGVtYV9kaV92aWRlb3NvcnZlZ2xpYW56YV9kZWxfY29tdW5lX2RpX21hY2VyYXRhLTk1MzU4MDguaHRt&f=Spages&file=L0ZJTEVTL0FsbGVnYXRpUGFnLzExOTMvUFJPVE9DT0xMT01BQ0VSQVRBMjNsdWcyMDIwLnBkZg==&id_sito=1193&s=download.php} \\
\scriptsize Memorandum of understanding & \scriptsize  Protocollo di intesa per la gestione del sistema integrato di videosorveglianza e di lettura targhe e dei comuni di Rovigo ed Occhiobello coordinato con i profili di interesse operativo delle forze di polizia territoriali. [English translation: Memorandum of understanding for the management of the integrated video surveillance and license plate reading system and the municipalities of Rovigo and Occhiobello coordinated with the operational interest profiles of the territorial police forces]. Available: \url{https://www.interno.gov.it/sites/default/files/2021-02/protocollo_videosorveglianza_rovigo_e_occhiobello_stralcio_funzionale_polesine_sicuro.pdf} \\
\scriptsize Online newspaper & \scriptsize  Videosorveglianza: Monteprandone aderisce ad un protocollo sovracomunale. Sarà realizzato un sistema centralizzato integrato in collaborazione con diversi Comuni, capofila Macerata. [English translation: Video surveillance: Monteprandone joins a supra-municipal protocol. A centralized integrated system will be implemented in collaboration with several municipalities, led by Macerata]. Available: \url{https://ascoli.cityrumors.it/prima-pagina/videosorveglianza-monteprandone-aderisce-ad-un-protocollo-sovracomunale-sara-realizzato-un-sistema-centralizzato-integrato-in-collaborazione-con-diversi-comuni-capofila-macerata.html} \\
\scriptsize Online newspaper & \scriptsize  Incontro a Macerata sull’istituzione di una rete di videosorveglianza sul territorio. [English translation: Meeting in Macerata on the establishment of a video surveillance network in the area]. Available: \url{https://www.maceratanotizie.it/40864/incontro-a-macerata-sullistituzione-di-una-rete-di-videosorveglianza-sul-territorio} \\
\scriptsize Online newspaper & \scriptsize  Macerata, nuovi sistemi di videosorveglianza: incontro tra i Comuni del territorio. [English translation: Macerata, new video surveillance systems: meeting between area municipalities]. Available: \url{https://picchionews.it/attualita/macerata-nuovi-sistemi-di-videosorveglianza-incontro-tra-i-comuni-del-territorio} \\
\scriptsize Memorandum of understanding & \scriptsize  Approvazione del "protocollo di intesa per la realizzazione di un sistema integrato di videosorveglianza (mediante l'utilizzo di telecamere OCR)". [English translation: Approval of the "memorandum of understanding for the implementation of an integrated video surveillance system (using OCR cameras)"]. Available: \url{http://www.comune.montaltodellemarche.ap.it/zf/index.php/atti-amministrativi/delibere/dettaglio/atto/G5WpRMETUWT0-A/provvedimenti/1} and \url{http://www.comune.montaltodellemarche.ap.it/c044032/de/attachment.php?serialDocumento=009NVR020202Q} \\
\hline

\bottomrule
\end{tabular}
\end{table*}

\vfill$\,$
\newpage

$\,$
\begin{table*}
\centering
\caption{Document' sources of DPO’s appointment}
\label{tab:DPO-appointment-documents}

\begin{tabular}{p{0.2\textwidth}p{0.7\textwidth}}
\toprule
Type & 
Document description (for URLs it refers to Section \ref{sec:DPOskills} of Table \ref{tab:other-case-studies-sources})\\
\midrule
City regulation & “Regolamento comunale per l’attuazione del Regolamento (UE) 2016/679 relativo alla protezione delle persone fisiche con riguardo al trattamento dei dati personali”, Tolentino, 2020. [English translation: Municipal Regulations for the Implementation of Regulation (EU) 2016/679 on the Protection of Individuals with regard to the Processing of Personal Data]. \\
County regulation & “Regolamento protezione dati personali della Provincia di Fermo,” Fermo, 2019. [English translation: Personal data protection regulations of the Province of Fermo]. \\
University regulation & Decreto Rettorale n. 137 del 17 aprile 2019, dell’Università degli Studi di Macerata [English translation:Rectoral Decree No. 137, April 17, 2019] \\
Ministry regulation & “Direttiva Ministro dello sviluppo economico in data 28 gennaio 2020 di individuazione dei soggetti attraverso i quali il Ministero dello sviluppo economico esercita le funzioni di titolare del trattamento ai sensi del Regolamento UE 2016/679”, Roma, 2020. [English translation: Directive Minister of Economic Development dated January 28, 2020 identifying the entities through which the Ministry of Economic Development exercises the functions of controller under EU Regulation 2016/679]. \\
Health Care Authority regulation & “Regolamento organizzativo aziendale privacy dell’Azienda sanitaria unica regionale delle Marche (Determina DG n. 349 del 30/05/2018)”, Ancona, 2018. [English translation: Corporate organizational privacy regulations of the Single Regional Health Authority of the Marche Region (DG Determination No. 349 of 30/05/2018)]. \\
Court Decision & “Sentenza del Tribunale Amministrativo del Friuli-Venezia Giulia, la n.287 del 13 settembre 2018”, Trieste, 2018. [English translation: Judgment of the Administrative Court of Friuli-Venezia Giulia, No. 287 of September 13, 2018]. \\
Call for an internal DPO’s appointment & “Bando di selezione interna per il conferimento dell’incarico di Responsabile della protezione dei dati (RPD) per l’Ateneo ai sensi dell’art. 37 del Regolamento UE 2016/679 (RGPD)”, Camerino, 2018. [English translation: Notice of internal selection for the appointment of Data Protection Officer (DPO) for the University in accordance with Article 37 of the EU Regulation 2016/679 (RGPD)]. \\
Call for an external DPO’s appointment & “Procedura negoziata in modalità telematica finalizzata alla esternalizzazione del servizio di D.P.O. (data protection officer) occorrente alla Azienda sanitaria unica delle Marche n. CIG 7409654FE6”, Ancona, 2018. [English translation: Negotiated procedure in telematic mode aimed at outsourcing the service of D.P.O. (data protection officer) needed by the Single Health Authority of the Marche Region]. \\
\bottomrule
\end{tabular}
\end{table*}

\vfill
$\,$

\begin{table*}
\centering 
\caption{Mapping Claims to Enforcement Tracker}
\label{tab:enforcement_tracker}
\footnotesize
\begin{tabular}{p{0.37\textwidth}p{0.08\textwidth}p{0.47\textwidth}}
\toprule
\textbf{GDPR's fines case types} & 
\textbf{Occurences} & 
\textbf{Reference scenarios} \\
\midrule

Insufficient fulfilment of information obligations & 124 & \WebsiteForcesChoices \\
Insufficient legal basis for data processing & 489 & \WebsiteForcesChoices, \WrongDPOAdvice, \AdminAsksForEverything, \WrongPublicProcurement \\
Insufficient fulfillment of data subjects rights & 136 & \NeglectedSubjectRight, \SubjectRightRequest \\
Insufficient technical and organizational measures to ensure information security & 274 & \WrongDPOAdvice, \DPOAdviceNotSought, \IgnoredDPOAdvice, \AdminAsksForEverything, \NoDataProtectionPrinciples, \UncheckedRemoteMonitoring, \WrongPublicProcurement, \SoftwareEndofLife \\
Non-compliance with general data processing principles & 332 & \IgnoredDPOAdvice, \WebsiteForcesChoices, \WrongDPOAdvice, \DPOAdviceNotSought, \AdminAsksForEverything, \UncheckedRemoteMonitoring, \WrongPublicProcurement, \SubcontractorViolatesPrivacy \\
Insufficient data processing agreement & 9 & \WebsiteForcesChoices, \WrongPublicProcurement \\
Insufficient fulfillment of data breach notification obligations & 25 & \NoDataProtectionPrinciples \\
Insufficient cooperation with supervisory authority & 56 & \NeglectedSubjectRight, \SubjectRightRequest \\
Insufficient involvement of data protection officer & 14 & \DPOAdviceNotSought \\
Unknown & 16 & \textit{These are particular cases that, according to GDPR Enforcement Tracker's site, are not generalizable.} \\

\bottomrule
\end{tabular}
\end{table*}

\clearpage

\section{Examples of DPO's functions}\label{sec:DPOtasks}

It is helpful to analyze from a practical point of view some scenarios in which the DPO carries out his or her tasks, to understand the seven functions of the DPO identified by EDPS. 

Typical examples of organizational function are high-risk processing. They will typically require the execution of a data protection impact assessment (DPIA) \cite{WP248r01}, which is a procedure that allows the controller to assess and demonstrate compliance with personal data protection laws. For example, in a local healthcare management organization, a DPO will help a controller to do a DPIA before starting new processing relating to a Covid-19 screening data acquisition and management system (Tab. \ref{tab:functionimpacted}, P2). Suppose the controller does not carry out this DPIA. In that case, a data breach may occur by exploiting an existing system vulnerability \cite{IlManifesto2021}; consequently, the competent supervisory authority (SA) sanctions the controller.
In another example, the DPO helps the controller assess risks posed by new processing built on an integrated territorial video surveillance system (implemented by OCR cameras). In that case, the DPO of the leading proponent municipality summons brings together and chairs a technical table with the mayors, the DPOs, and the technical delegates of all the Municipalities involved (this is a case of joint controllers). This committee defines all the project's technical and organizational aspects (Tab. \ref{tab:functionimpacted}, P1). According to the privacy by design principle, this assessment must happen before starting the processing activities. After that, if the analysis results confirm that the processing complies with the data protection laws, the DPO updates the records of processing activities. An example of monitoring activity is personal data breach management. (Tab. \ref{tab:functionimpacted}, R2, R7, and R10). An example of monitoring activity is personal data breach management. One more example of this function is the internal complaints handling the investigation. For example, an employee reports to the DPO that the staff responsible for checking the possession of the EU Digital COVID Certificate for access to the workplace illegally also controls the specific category of the certification to understand which persons are not vaccinated against COVID-19. Upon this notification, the DPO initiates an investigative activity to verify the correctness of the control procedures of the EU Digital COVID Certificate held by the staff (Tab. \ref{tab:functionimpacted}, I3).

Another function is supporting the controller to comply with data protection by design principle. In Tab. \ref{tab:functionimpacted}, D1 and D2, we described further examples. In another typical example, after a personal data breach occurs, the DPO of a Health Care Authority summons brings together and chairs a technical table with the area managers. Then he or she reports directly to the controller, and if a breach results in a high risk to the rights and freedom of natural persons, communicate the breach to data subjects on behalf of the controller (Tab. \ref{tab:functionimpacted}, R10). A similar case is reported in Tab. \ref{tab:functionimpacted}, R7.

For the cooperation function, the DPO helps the SA facilitate access to the documents and information both for the performance of the SA's tasks and for the exercise of the latter's investigative, corrective, authorization, and advisory powers (Tab. \ref{tab:functionimpacted}, R8). By carrying out these tasks, the DPO \textbf{is not} an agent of the SA; he or she acts as an expert that is part of the controller's organization in which he or she works.

Relating to the handle queries or complaints function, the DPO must give a prompt response to data subjects who exercises their rights (Tab. \ref{tab:functionimpacted}, R11 and R12).

For the information and awareness-raising function in small and medium-sized enterprises (SME), first, DPOs monitor both the data protection laws changes and the indications of the bodies in charge, and then they prepare short information pills for the staff or create some quick information notes for the team. These informational materials could be made available also by publishing them on a specific internal website (Tab. \ref{tab:functionimpacted}, P4).

Despite the DPO having the competence to monitor compliance with the GDPR and handle complaints, his enforcement powers are minimal. For example, in the case of new processing that could present a high risk for the rights and freedoms of natural persons, the controller consults with the DPO on whether or not to carry out a DPIA. If a DPIA is necessary, this consults about the best methodology to follow when carrying out that DPIA and which safeguards to apply to mitigate any related risks. If the controller disagrees with the advice provided by the DPO, it may not take it into account; the controller is only obliged to document in writing why it chose to operate like this. Indeed, according to the accountability principle, the advice of a DPO is not binding for a controller (Tab. \ref{tab:functionimpacted}, R1, R3, and R5).

\clearpage

\section{A Map of Risks}\label{sec:DPOchallenges}
The challenges of DPOs are varied, as the organizations they work for. However, it is helpful to classify them into two categories of risks that are typical of a socio-technical system:
\begin{itemize}
    \item \textbf{Technical Risks (TR):} resulting from technical or technological solutions implemented in processing that does not guarantee the protection of the subjects whose personal data is processed.
    \item \textbf{Socio-technical Risks (SR)}: resulting from not optimal execution or designing of organizational procedures, as well as from an incorrect interaction between human actors.
\end{itemize}

\subsection{Technical Risks (TR)}\label{sec:TV}
For the readers of this magazine, Technical Risks are likely the most interesting. We can subdivide them into two other groups. The first type relates to design problems, such as choosing a configuration that complies with the data protection by default principle or choosing insecure technical solutions in the call for tenders. The second type involves misconfigurations or errors which appear during the implementation.

The DPO must support and advise the controller on many technical choices. 
For example, the choice of technical and organizational measures ensures that, by default, only personal data necessary for each specific purpose are processed. The incorrect technical configuration of a telecommunication software switchboard may result in a data breach (Tab. \ref{tab:scenarios}, \NoDataProtectionPrinciples) because everyone can view and download the call's metadata and registration. In another example (\UncheckedRemoteMonitoring), an incorrect processing design may cause choosing tools that do not use encryption or allow unchecked remote desktop connections.

A further example is related to the selection procedures for technical solutions in calls for tenders (\WrongPublicProcurement) if applicants are not required to “demonstrate” the GDPR compliance of their products or services. As a result, applicants that can demonstrate compliance with the GDPR do not receive a competitive advantage compared to the others. The tender commission may not be able to motivate the choice of the winner of the tender technically. Worse, this omission could make a compliant contractor lose a bid against a not-compliant contractor and kick-start litigation. In the scenario (\SubcontractorViolatesPrivacy), a tender winner is violating the contract, for example, because the processor owns the data encryption key and could read all the personal data processed.

In \SoftwareEndofLife, there is an example of a technical choice for a company that uses obsolete software. In this case, a DPO noticed that many workstations used a version of the Microsoft Windows operating system, obsolete or with the vendor’s extended support expired. 

\subsection{Socio-technical Risks (SR)}\label{sec:DFV}
Socio-technical risks can appear daily while the DPO supports the controller in arranging processing activities.

We can specialize these risks into four additional categories, such as \emph{auditing} (e.g., Tab. \ref{tab:scenarios}, \IgnoredDPOAdvice\ and \SubcontractorViolatesPrivacy\ ), \emph{communication} (e.g., \NoDataProtectionPrinciples\ and \UncheckedRemoteMonitoring ), \emph{processing designing} (e.g., \SoftwareEndofLife ) and \emph{relationship} (e.g., \NeglectedSubjectRight ).

The first is related to activities in which the DPO works on the processing inventory or processing auditing, including keeping a record of the processing activities and the risk analysis or the DPIA execution. The second involves all the issues in which we have an exchange of information between two distinct entities, for example, between the DPO and the staff involved in the processing activities or between the DPO and the controller's management. The third is related to activities of new processing design or an existing modification. The last refers to relationships between the controller and other different bodies (such as a processor, a third party, a data subject, or a SA).

Conflicting requirements are a particular instance of these risks. An example is the conflicting data flow resulting from the execution of procedures meeting conflicting requirements (e.g., freedom of information acts vs. data protection acts) where personal data of the involved subjects may (accidentally or maliciously) surface.

DPOs often experience \emph{side-channel attacks} from the most unexpected sources.
A well-designed processing activity may become unlawful because a staff member operates differently from what is prescribed by the procedure and from the instructions received by the controller.

For example, a DPO discovers that a local online newspaper released a video of a traffic accident, which appears to derive from some closed-circuit video surveillance cameras used by the local police. In that case, the risk lies in something other than the failure of data flow management or the choice of technical solutions. Rather presumably, it lies in the unplanned wickedness of an operator who had chosen to film the screen of the closed-circuit control station with the smartphone and disseminate the video on social channels until it was re-broadcasted by the local newspaper.

In another example, a gym employee registers a new customer who signs up for a quarterly access subscription, asking her to sign her consent to processing personal data for marketing purposes. Unfortunately, violating the organization's procedures could expose the controller to administrative fines or penalties if the employee takes this consensus without first delivering a copy of the information to the data subject.

\end{document}


\title{Additional material number 1}

\maketitle

To understand what the DPO role concretely means it is useful taking into account the seven functions identified by the EDPS \cite{EDPS2005} \cite{Korff2019}, as listed in Table \ref{tab:function}.

A Typical example of the organizational function is the creation of a register of personal data processing operations, the consequential review of the personal data processing operations, the assessment of the risks posed by these, and finally the subsequent management of those operations that have a high correlated risk value. A high risk processing will typically require the execution of a DPIA (data protection impact assessment), that is a procedure that allows the controller to assess and demonstrate compliance with the personal data protection laws. For example, in a local health care management organization a DPO will help a controller to do a data protection impact assessment before starting a new processing relating to a Covid-19 screening data acquisition and management system. This is a particular process that satisfies at least two of the criteria indicated by the old Article 29 Working Party independent European advisory body, in its guidelines related to the data protection impact assessment execution \cite{WP248r01}. If this DPIA is not carried out, it is possible that a data breach occurs exploiting an existing vulnerability in the system \cite{IlManifesto2021}, and that consequently the competent supervisory authority sanctions the controller \cite{Garante09/2022}.

Another example, according to privacy by design principle, prior to starting new processing built on an integrated territorial video surveillance system (implemented by using OCR cameras), the DPO helps the controller to assess the risks posed by this processing. After that, if the analysis' results confirm that the processing is compliant with the data protection laws, the DPO updates the register of personal data. Example monitoring activity is personal data breaches management. The DPO, summons, brings together and chairs a technical committee with the area managers to investigate this topic, then reports directly to the controller for subsequent actions to be taken. One more example of this function is the internal complaints handling investigation.  For example, an employee reports to the DPO that the staff responsible for checking the possession of EU Digital COVID Certificate (\cite{EUCovidCertificate},  \cite{RegulationEU2021/953}, \cite{DL52/2021}, and subsequent amendments and additions) for access to the workplace, illegally controls also the specific category of the certification in order to understand which persons are not vaccinated against COVID-19. Upon this notification, the DPO initiates an investigative activity to verify the correctness of the control procedures of the EU Digital COVID Certificate held by the staff.

Another function is the advisory one, related to the more concrete protection by design principles. For example, according to privacy by design principle, a consortium of different municipalities informs their DPOs that they want to start a new processing built on an integrated territorial video surveillance system (using OCR cameras). The DPO of the leading proponent municipality summons, brings together and chairs a technical table with both the mayors and the DPOs and the technical delegates of all the Municipalities involved, defining all the project's technical and organizational aspects. Finally, this working group preliminarily arranges a protocol of understanding to be signed by other municipalities. In this document there is a formal undertaking to jointly proceed with the prior consultation of the supervisor authority by DPOs. A further common example is that in which the DPO of a Bank is invited to participate in periodical meetings of Area Managers, especially when issues with data protection implications are discussed. In another typical example, after a personal data breach occurred, the DPO of an Health Care Authority summons, brings together and chairs a technical with the area managers, then he reports directly to the controller, and in case of a breach resulting in high risk to rights and freedom of natural persons, communicates, on behalf of controller, the breach to the data subjects \cite{ASURMarche2018-RO}. In another case, the DPO of a ministry advises the Administration that in issuing public tenders it should expressly call for applicants that can “\emph{demonstrate}” that their product or service fully complies with any relevant EU and national data protection law. For this purpose the DPO suggests that if an applicant can demonstrate this compliance it should receive a competitive advantage.

From a practical point of view, for the cooperation function the DPO helps the supervisory authority facilitating access to the documents and information both for the performance of the DPA's tasks, and for the exercise of the latter's investigative, corrective, authorisation, and advisory powers. By carrying out these tasks the DPO \textbf{is not} an agent of the supervisory authority, in fact, it acts as an expert that is part of the controller's organization in which he works. 

To handle queries or complaints function, the DPO must monitor his or her contact addresses (such as email or telephone number) and provide a prompt response (without undue delay, at the latest within 1 month of receipt) to the request. For example, the DPO of a Municipality, after receiving a telephone request from a data subject, provides him with information on a processing, also indicating the exacts URLs of the web site of the controller from which he can download both a copy of the information about the personal data process, and a specific application form, useful for present a motion for the exercise of the rights provided by articles 15-22 of the GDPR \cite{GDPR}.

For the information and awareness raising function, in a small and medium-sized enterprise (SME) first a DPO monitors both the data protection laws changes and the indications of the bodies in charge, and then he prepares short information pills for the staff, or creates some short information notes for the staff. These informational materials could be made available also by publishing it in a specific internal web site. The DPO of a Bank illustrates how in the reference year it advised the controller about their obligations pursuant to the revision of the internal regulations on conflict of interest, or the treatment of cases of “homocody” (coincidence of tax codes) in the Central Credit Register \cite{BancadItalia2020}.

Despite the DPO having competence both to monitor compliance with the GDPR \cite{GDPR} and to handle complaints; his powers of enforcement are very limited. For example, in the case of a new processing that could present a high risk for the rights and freedoms of natural persons, the controller consults with the DPO on whether or not to carry out a DPIA and, if so, about which is the best methodology to follow when carrying out a DPIA, and which are the need safeguards to apply to mitigate any related risks. If the controller disagrees with the advice provided by the DPO, he may not take it into account; in fact the controller is only obliged to document in writing why it chose to operate like this. Indeed, according to the accountability principle, the advice of a DPO is not binding for a controller.

However, the edge edge of the competence of these functions sometimes is fuzzy, specially in complex cases studies where more of them are involved. In Table \ref{tab:functionimpacted}, we summarized some examples of these cases.

\bibliographystyle{IEEEtran}
\bibliography{dpobib}


\title{Additional material number 2}

\maketitle

To understand what the DPO role concretely means it is useful taking into account the seven functions identified by the EDPS \cite{EDPS2005} \cite{Korff2019}, as listed in Table \ref{tab:function}.
\def\High{$\star\star\star$}
\def\Medium{$\star\star$}
\def\Low{$\star$}

Table \ref{tab:vulnerabilities} summarises in a matrix notation how much each of the seven main functions of the DPO is affected by each of the vulnerability categories described in the article. In the Table \ref{tab:vulnerabilities} the coding used is as follows: "\emph{-}" if the function is not affected by the vulnerability, otherwise the number of stars (\Low) to specify the grade with which the DPO function is involved from the vulnerabilities or rather the extent to which the presence of the vulnerability in question could make carrying out each function (in detail, \Low\ for a low value, \Medium\ for a medium value, and \High\ for an high value).

\begin{table}
\centering
\caption{DPO Vulnerabilities}\label{tab:vulnerabilities}
\begin{minipage}{\textwidth}\footnotesize
This table describes some different socio-technical vulnerability categories that could affect the DPO's functions 
\end{minipage}
\begin{tabular}{p{2cm}p{2.5cm}p{1.5cm}p{1.5cm}p{1.5cm}p{1.5cm}p{1.5cm}p{1.5cm}p{1.5cm}}
\toprule
\textbf{Vuln. category} & 
\textbf{Vulnerability type} & \textbf{Organisational function} & \textbf{Monitoring of compliance} & \textbf{Advisory function} & \textbf{Cooperative function} & \textbf{Handle queries or complaints} & \textbf{Information and raising awareness} & \textbf{Enforcement} \\
\midrule

\multirow{4}{2.5cm}{\textbf{Conflicting Data Flow}} &  Auditing & \High & \High & \High & \Low & \Medium & \Low & \High \\
 & Communication & \Medium & \High & \High & \Medium & \High & \High & \High \\
 & Process designing & \High & \Medium & \High & \Medium & - & \Medium & \High \\
 & Relationship & \Medium & \High & \Medium & \High & \High & \Medium & \Medium \\
\hline

\multirow{2}{2.5cm}{\textbf{Technical Issues}} & Wrong Design & \High & \Medium & \High & \Medium & - & \Medium & \High \\
 & Misconfiguration & \Medium & \High & \Medium & \Low & \Low & \Low & \Low \\
\bottomrule
\end{tabular}
\end{table}

\bibliographystyle{IEEEtran}
\bibliography{dpobib}


\title{Additional material number 4 - (Methodology)}

\maketitle

\section{Methodology short description} \label{AM4:Methodology}
In this article we used a methodology to derive the scenario from some sources of information publicly available, without basing our consideration only on the subjective perception of one of the authors.

Hence, the insights described in this article are derived using the following listed methodology, and summarized in the Diagram \ref{fig:scenarios_methodology}, which illustrates the steps followed to construct the reference scenarios analyzed in the article.
\begin{enumerate}
    \item Analyzing the existing data protection laws as well as opinions and recommendations of relevant authorities (such as the EDPS and the EDPB).
    \item Analyzing the seven DPO’s functions that the European Data Protection Supervisor identified in its position paper on the role of DPO in compliance with Regulation (EC) 45/2001.
    \item Identifying all the DPO’s activities which are involved to the DPO's functions, and correlating them together. In making this position, we take into account how some EU SAs (namely the supervisory authorities of Bulgaria, Croatia, Italy, Poland and Spain, which are involved in the T4DATA EU project) used these functions to group all the DPO’s tasks. This grouping is described by D. Korff and M. Georges in “The DPO Handbook Guidance for data protection officers in the public and quasi-public sectors on how to ensure compliance with the European Union General Data Protection Regulation”.
    \item Using of personal experience to parse the case studies related to the DPO's tasks, by selecting the more illustrative, as described in deep in the next step.
    \item Identified some illustrative case studies of the DPO’s tasks in which the activities (identified in the previous steps) are carried out.
   \item Looking for some publicly available information and used them as reference to illustrate the case studies.
   \item Masking the selected publicly available information (but without altering their essence) in order to become the actors who are involved not directly recognizable.
    \item Building the article's scenarios. Using the case studies, we carried out analytical work on these to identify possible vulnerabilities and the possible consequences associated.
\end{enumerate}

\begin{figure}[h]
\centering
\captionsetup{justification=centering} 
\includegraphics[scale=.55]{2022_08_22_Methodology_diagram.png}
\caption{Methodology's diagram}
\label{fig:scenarios_methodology}
\end{figure}

\section{Scenarios' sources evidence}\label{AM4:Sources}
The case studies analyzed in our work are those of the final scenarios building in step 8 of Section \ref{AM4:Methodology}. In building our scenarios we followed the methodology suggested by Yin in \cite{yin2018case}. According to this vision, despite one of the most important sources of case study evidence being the interview, this is not the only one. In particular, there could be multiple (at least six) sources of evidence for a case study building. 
Hence the better case studies rely on a variety of sources that usually are used together to build triangulation. In our work we used multiple sources of evidence that have been processed according to the methodology described in Diagram \ref{fig:scenarios_methodology}. Direct observation, documentation and archival records are used in step 4 and 5 using the first author's personal experience to identify some intermediate case studies related to the DPO’s tasks that are particularly illustrative of the activities carried out by this role. Instead, documentation and archival records (which are publicly available) are used in step 6 as a reference to illustrate with publicly available evidence the previous intermediate case studies. 
Prior to building the final scenarios (which are the final case studies analyzed in our work) these evidences are masking according to the step 7 content.

\section{Where the scenarios come from?} \label{AM4:Scenarios}
The scenarios come from the personal experience of the first author, but to illustrate them we took into account a series of cases that are in the public domain (for example, judgments, revert in court, supervisory authorities’ decisions, newspaper articles). The scenarios were chosen so we can have at least one analyzable practical example for each of the DPO's main tasks, as they are defined by the relevant authorities. We can make the list of the original source material as an additional material, as better detailed replies to document named \textit{Additional material number 3}.

The authors do not directly cited them for the following two reasons:
\begin{itemize}
    \item there is a limitation in the number of citations usable in the article (it must be not more than 15), while the number of case studies is wide larger;
    \item they choose to focus on the scenarios characteristics, avoiding making them finger-printable, in order to minimize the possibility that in the future the scenarios’ actors can exercise their right to be forgotten.
\end{itemize}
For example, it is possible (because it has happened in the past) that in the future a judgment  will force a newspaper to remove some personal data mentioned in a previously published article. Hence, because we masked they, the Magazine is safe from these kinds of problems.

\bibliographystyle{IEEEtran}


\bibliography{dpobib}